\documentclass[galaxies,article,accept,pdftex,moreauthors]{Definitions/mdpi}
\usepackage{xspace}
\usepackage{aas_macros}

\firstpage{1}
\pubvolume{1}
\issuenum{1}
\articlenumber{0}
\pubyear{2023}
\copyrightyear{2022}
\externaleditor{Academic Editor: Stefi Baum, Christopher P. O'Dea} 
\datereceived{9 January 2023} 
\daterevised{20 March 2023}
\dateaccepted{28 March 2023} 
\datepublished{6 April 2023} 
\hreflink{https://doi.org/}

\usepackage{xcolor}
\usepackage[normalem]{ulem}
\usepackage{footmisc}

\newcommand{\Msun}{\ifmmode {$M$_{\odot}}\else{M$_{\odot}$}\fi}

\Title{Intermediate-Mass Black Holes: The Essential Population to Explore the Unified Model for Accretion and Ejection Processes}

\TitleCitation{Intermediate-Mass Black Holes: The~Essential Population to Explore the Unified Model for Accretion and Ejection Processes}

\Author{Xiaolong Yang{}$^{1,2,}$*\orcidA{}, Jun Yang $^{3}$\orcidB{}}

\AuthorNames{Xiaolong Yang and Jun Yang}

\AuthorCitation{Yang, X.; Yang, J.}

\address[1]{$^{1}$ \quad Shanghai Astronomical Observatory, Chinese Academy of Sciences, Shanghai 200030, China\\
$^{2}$ \quad Shanghai Key Laboratory of Space Navigation and Positioning Techniques, Shanghai 200030, China \\
$^{3}$ \quad Department of Space, Earth and Environment, Chalmers University of Technology, Onsala Space Observatory, SE-439 92 Onsala, Sweden}

\corres{\hangafter=1 \hangindent=1.05em \hspace{-0.82em} Correspondence: yangxl@shao.ac.cn}

\abstract{We study radio and X-ray emissions from intermediate-mass black holes (IMBHs) and explore the unified model for accretion and ejection processes. The radio band survey of IMBH (candidate) hosted galaxies indicates that only a small fraction ($\sim$0.6\%) of them are radio-band active. In addition, very long baseline interferometry observations reveal parsec-scale radio emission of IMBHs,  further resulting in a lower fraction of actively ejecting objects (radio emission is produced by IMBHs other than hosts), which is consistent with a long quiescent state in the evolution cycle of IMBHs. Most (75\%, i.e., 3 out of 4 samples according to a recent mini-survey) of the radio-emitting IMBHs are associated with radio relics and there is also evidence of dual radio blobs from episodic ejecting phases. Taking the radio emission and the corresponding core X-ray emission of IMBH, we confirm a universal fundamental plane relation (FMP) of black hole activity. Furthermore, state transitions can be inferred by comparing a few cases in XRBs and IMBHs in FMP, i.e., both radio luminosity and emission regions evolve along these state transitions. These signatures and evidence suggest an analogy among all kinds of accretion systems which span from stellar mass to supermassive black holes, hinting at unified accretion and ejection physics. To validate the unified model, we explore the correlation between the scale of outflows (corresponding to ejection powers) and the masses of central engines; it shows that the largest scale of outflows $\hat{LS}_\mathrm{out}$ follows a power-law correlation with the masses of accretors $M_\mathrm{core}$, i.e., $\log{\hat{LS}_\mathrm{out}} = (0.73\pm0.01)\log{M_\mathrm{core}} - (3.34\pm0.10)$. In conclusion, this work provides evidence to support the claim that the ejection (and accretion) process behaves as scale-invariant and their power is regulated by the masses of accretors.}


\begin{document}
\section{Introduction}
Two types of astrophysical black holes, stellar-mass black holes (SBHs) and supermassive black holes (SMBHs), are widely identified in the Universe through their observational signatures. The~death of massive stars will leave behind SBHs as fossil~\citep{2017NewAR..78....1M}, which have the masses 3--100 $M_\odot$. SBHs are abundant in our Galaxy as the X-ray binary systems (XRBs, which accrete matters from a companion star). On~the other hand, SMBHs \mbox{($\geq$$10^6$ $M_\odot$)} are generally identified at the center of massive galaxies (they co-evolve with bulges and hosts)~\citep{2013ARA&A..51..511K}. Remarkably, SMBHs have an essential role in regulating the evolution of their host galaxies, which conduce to the correlation between SMBHs and their host properties, e.g.,~the correlation between the SMBH mass and the bulge velocity dispersion ($M-\sigma$). In particular, accreting SMBHs from active galactic nuclei (AGNs) can maintain strong feedback to the hosts. The~black holes with masses between the stellar mass black holes and the supermassive black holes are the so-called intermediate-mass black holes (IMBHs, $10^2$--$10^6\, M_\odot$, see the review by~\citep{2020ARA&A..58..257G}).

We now come to realize that the existence (or not) of IMBHs is a fundamental question; it plays an important role in the supermassive black hole formation and evolution over the cosmic time (see~\citep{2020ARA&A..58..257G} and references therein). The~discoveries of SMBHs with masses up to $10^{10}\,M_\odot$~\citep{2015Natur.518..512W, 2018Natur.553..473B} in the early Universe, when the Universe was only $5\%$ of its current age, pose challenges to the formation of SMBHs. Mergers and accretions are the two reasonable channels of growth to such high masses (see~\citep{2010A&ARv..18..279V,2020ARA&A..58..257G}). Growing up via stellar-mass seed black holes would require extremely high accretion rates (e.g.,  super-Eddington accretion, Eddington ratio $\lambda_{\rm Edd} = L_{\rm bol}/L_{\rm Edd} \gtrsim 1$, where $L_{\rm bol}$ is the bolometric luminosity and $L_{\rm Edd}$ is the Eddington luminosity); however, this short-lived accretion state seems inefficient in affording the mass explosion~\citep{2020ARA&A..58...27I}. On~the other hand, if~growing up from the intermediate-mass seed black holes, then the accretion process would be more effective with respect to assembling SMBHs in the early Universe (see~\citep{2010A&ARv..18..279V,2017IJMPD..2630021M,2020ARA&A..58..257G,2020ARA&A..58...27I} and references therein). However, the~relatively smaller mass (than SMBHs) and greater distance (than Galactic SBHs) of IMBHs make them difficult to discover in observations (especially those with masses $10^3$--$10^4\, M_\odot$).

Several methods can be hired to constrain the mass of IMBH candidates, for~example, the~correlation between the black hole mass and bulge velocity dispersion (e.g.,~\citep{2013ARA&A..51..511K,2019MNRAS.484..794G,2019MNRAS.484..814G}), the~fundamental plane of black hole activity (e.g.,~\citep{2012ApJ...753...38S,2015MNRAS.448.1893M,2022MNRAS.517.4959Y}), the~reverberation mapping (e.g.,~\citep{2000ApJ...533..631K,2019NatAs...3..755W}, modeling the light curves (e.g.,~\citep{2012ApJ...745...89L,2021NatAs...5..560P}), the~virial mass based on the broad H$\alpha$ line~\citep{2004ApJ...610..722G, 2007ApJ...670...92G, 2012ApJ...755..167D, 2013ApJ...775..116R} and the inverse correlation between the black hole mass and the high-frequency quasi-periodic oscillations~\citep{2004ApJ...609L..63A,2012A&A...548A...3A,2014Natur.513...74P,2015ApJ...811L..11P,2019A&A...622L...8G}. In particular, a~consistent black hole mass from multi-method can be considered as the best-constrained value; for~instance,~\mbox{\citet{2014Natur.513...74P}} estimated the black hole mass of M82 X-1 as $\sim$400\Msun through both quasi-periodic oscillations and spectral variability methods. The searches for IMBHs in observations, according to the formation and evolution scenarios, are typically focused on several habitats, e.g.,~ultra-/hyper-luminous X-ray sources (e.g.,~\citep{2004MNRAS.355..413P, 2004ApJ...609L..63A, 2012Sci...337..554W, 2014Natur.513...74P, 2015ApJ...811L..11P, 2017ARA&A..55..303K}), globular clusters (e.g.,~\mbox{\citep{2005ApJ...634.1093G, 2017Natur.542..203K, 2021ApJ...918...46W, 2022ApJ...924...48P}}) and~dwarf galaxies (e.g.,~\citep{2003ApJ...588L..13F, 2004ApJ...610..722G}). Among~them, dwarf galaxies with total stellar masses ranging from $10^7$--$10^{10}\,M_\odot$ (e.g.,~\citep{2004ApJ...610..722G, 2007ApJ...670...92G, 2012ApJ...755..167D, 2018ApJS..235...40L, 2018ApJ...863....1C}) tend to contain active IMBHs. Moreover, dwarf galaxies and their central intermediate-mass (typically $10^4$--$10^6\,M_\odot$) black holes are thought to have already appeared in the early Universe~\citep{2010A&ARv..18..279V,2012NatCo...3.1304G}. The~dwarf galaxies which live in the present day have to evolve in isolation (i.e., undergo a few mergers); therefore, their central black holes are still in the baby stage and provide clues for the formation and evolution of SMBHs from the IMBH seeds~\citep{2010A&ARv..18..279V,2020ARA&A..58...27I}.

The study of IMBHs is essential for exploring the unified accretion and ejection model spanning from stellar to supermassive black holes~\citep{1995A&A...293..665F,2003MNRAS.345.1057M, 2014ApJ...788L..22G}. Accretion and ejection are the two important and universal phenomena of compact objects. The~X-ray and radio emissions radiated from the central engines of the stellar mass accreting X-ray binary systems are thought to be primarily and generally from accretion disks (and coronae) and collimated jets, respectively, which provide an observational avenue to investigate the disk and jet correlation with the black hole mass~\citep{2003MNRAS.345.1057M}. Due to the low radio-loud fraction of low-mass AGNs with high accretion rates (see the pilot radio survey and study by \citet{2006ApJ...636...56G}), one draws a comparison between the low-mass AGNs and Galactic X-ray binaries: the radio emissions in the high accretion rate and soft X-ray (high/soft) state of XRBs are suppressed (e.g.,~\citep{2006csxs.book..157M}), similar to the radio properties observed in the low-mass and high accretion rates AGNs. As~a reasonable hypothesis, the~physical process in both XRBs and AGNs is similar and scale-invariant, then the observed radio properties can be well characterized in terms of accretion states and transition. Further evidence comes from a recent study by \citet{2020ApJ...904..200Y}, who studied a sample of AGNs with extremely high accretion rates (near or above the Eddington limit) and found an inverse correlation between the radio loudness $\mathcal{R}$ and the Eddington ratio $\lambda_{\rm Edd}$; the~results indicate that AGNs accreting at high and super-Eddington rates are predominately~radio-quiet.

The analogy between XRBs and AGNs implies a scale-invariant accretion and ejection process among accreting black holes. This is motivated by a radio-to-X-ray correlation (i.e., the disk--jet coupling) in the XRB systems~\citep{2003MNRAS.344...60G}; then \citet{2003MNRAS.345.1057M} shows that the radio-to-X-ray correlation is applicable in both XRBs and AGNs and is regulated by the black hole masses, i.e.,~the fundamental plane of black hole activity. Unfortunately, the~original 'fundamental plane' has a break between SBHs and SMBHs due to the lack of IMBH samples. Subsequent studies, e.g., \citet{2006ApJ...636...56G}, \citet{2014ApJ...788L..22G} and~\mbox{\citet{2022MNRAS.513.4673B}} investigated the radio properties of the low-mass RQ-AGNs using the arcsec-scale resolution Very Large Array (VLA) observations and found that they deviate from  Merloni's fundamental plane relation. However, our recent studies~\citep{yang2022radio} show that the inconsistency between the fundamental plane and the observations for low-mass AGNs may simply correspond to a deficit of~resolution.

Radio emissions from radio-loud AGNs (RL-AGNs; defined as $\mathcal{R} > 10$) are primarily from synchrotron mechanisms, where the electrons are accelerated to relativistic energies in jets (e.g.,~\cite{1979ApJ...232...34B,2019ARA&A..57..467B}). Jets can exist for a long time (up to mega years) and the physical scales of the relativistic jets can extend from less than a parsec to megaparsec scales. Jets are the dominating radio-emitting source in RL-AGNs; however, most AGNs are radio-quiet~\citep{2015ApJ...804..118B}. On~the weak radio emissions of radio-quiet AGNs, several mechanisms are responsible for the radio origin (see~\cite{2019NatAs...3..387P} and references therein). The~possible radio emissions of RQ-AGNs include lower power jets (e.g.,~\cite{2020MNRAS.494.1744Y,2021MNRAS.500.2620Y,2021MNRAS.504.3823W}), free-free emission from photo-ionized gas in the circum-nuclear region~\citep{2021MNRAS.508..680B}, star-forming regions (e.g.,~\cite{2021MNRAS.500.2620Y}), accretion disk--corona activity (e.g.,~\cite{2008MNRAS.390..847L, 2016MNRAS.459.2082R,2018ApJ...869..114I,2020ApJ...904..200Y}) and sub-relativistic wide-angled winds (e.g.,~\cite{2014MNRAS.442..784Z, 2015MNRAS.447.3612N,2021MNRAS.504.3823W}). Hereinafter, we term the jets and the wind-like ejections as outflows. Therefore, in~constraining the radio emission of radio-quiet AGNs, high-resolution (e.g., the very long baseline interferometry) radio observations are~essential.

As one of the highest-resolution approaches, the~very long baseline interferometry (VLBI) method at the radio band provides the milli-arcsec scale resolutions at GHz bands. The~high-resolution VLBI observations of dwarf galaxies can reveal pc-scale radio-emitting structures of their central engines, which can be cores, core-jets or~jet-knots. Therefore, VLBI observations can directly probe the jet powered by potentially accreting IMBHs. At~present, the~VLBI observations of IMBH candidates in dwarf galaxies are only limited to a few cases: NGC~4293~\citep{2022MNRAS.517.4959Y}, NGC~4395~\citep{2006ApJ...646L..95W}, Henize~2-10~\citep{2012ApJ...750L..24R}, NGC~404~\citep{2014ApJ...791....2P}, RGG~9~\citep{2020MNRAS.495L..71Y} and the four samples from the mini-survey of IMBHs by \citet{yang2022radio}. The~radio observation with the High Sensitivity Array (HSA) 1.4\,GHz of NGC~4395 ($R_{\rm X} = -5$, ref.~\cite{2003ApJ...583..145T}) reveals an elongated sub-parsec scale structure, which was explained as a core-jet~\citep{2006ApJ...646L..95W}; however, the~recent study with European VLBI Network (EVN) 5 GHz high-resolution observations of NGC~4395 by \citet{2022MNRAS.514.6215Y} failed to detect a radio-active nucleus at the optical ({\it Gaia}) position. Alternatively, a~relatively low-resolution Very Large Array (VLA) 15\,GHz observation detected the radio emission that coincides with the {\it Gaia} position. The~above observations are consistent with the scenario of multiple radio lobes from episodic ejections. On~the other hand, the~EVN 5\,GHz observation of NGC\,404~\citep{2014ApJ...791....2P} failed to detect radio emission from its nucleus, while the large-scale (parsec-scale) radio emission was detected with the VLA 1.4 GHz observations~\citep{2012ApJ...753..103N}. Furthermore, the~{\it Chandra} X-ray observation of NGC\,404 indicated a steady source of $\log{R_\mathrm{X}}<-3.8$ in the~nucleus.

In this work, we explore the universal accretion and ejection processes spanning from stellar mass to supermassive black holes. In particular, this work involves the largest sample (at present) of radio-emitting IMBHs to unveil the vacuum region between SBHs and SMBHs. Throughout this work, we adopt the standard $\Lambda$CDM cosmology with a Hubble constant $H_0 =70$ km s$^{-1}$ Mpc$^{-1}$ and~matter density and dark energy density parameters $\Omega_\Lambda=0.73$, $\Omega_m=0.27$, respectively.

\section{The Universal Picture for Black Hole Accretion and~Ejection}
Both SBHs and SMBHs can maintain accretion and ejection processes, which form X-ray binaries (XRBs) and active galactic nuclei (AGNs), respectively. The~accretion discs and associated coronae and jets in both XRBs and AGNs produce multi-band emissions, especially at X-ray and radio bands. Though~with different environments and size scales, observations and theoretical studies reveal similar accretion and ejection behaviors, e.g.,~the evolution of accretion flows, the~structures of accretion discs and~the features of the ejection process. In~individual sources, the~accretion and ejection processes evolve as the fluctuation of accretion rates in terms of Eddington ratios (see~\citep{1997ApJ...489..865E, 2017SSRv..207....5R}).

On the activity of SBHs, monitoring and simultaneous observations at radio and X-ray bands reveal a clear evolution cycle among them. They can be well characterized as several accretion and ejection states through X-ray and radio properties~\cite{2004MNRAS.355.1105F}. The~X-ray emissions in accreting systems are thought to be primarily and generally from the accretion disk and corona system (and jets as well), while radio emissions are thought to be dominated by jets. The~association between X-ray and radio properties in X-ray binary indicates a strong coupling between accretion discs and jets~\cite{1997ApJ...489..865E}. The~jet--disk coupling in X-ray binaries can be observed with convenient time scales (months to years), which is long expected to be a common feature in both stellar mass and supermassive black holes. Furthermore, observations of X-ray binaries show that the accretion flow and jet production is correlated with the accretion rate/Eddington ratio. The~fluctuations of the accretion rate will essentially trigger the accretion state transition~\citep{2019ApJ...883...76R}. 

Based on the monitoring observations of X-ray binaries, we now have a picture of state transition and evolution: (a) the accretion with a low Eddington ratio (typically below a few percent of the Eddington luminosity) is characterized by a `low/hard' X-ray state; it is associated with compact radio emission and identified to be steady and short jets; (b)~with the increase of accretion rates, the~X-ray spectrum becomes dominated by the soft emission and the radio emission dramatically drops to an undetectable level. The~Eddington ratio of the soft state is substantially close to the Eddington limit; (c) the transition from the `low/hard' state to the `soft state' is generally associated with an unstable `very high state', where the Eddington ratio will be up to and over one, i.e.,~near- and super-Eddington rates. Episodic jet blobs can be triggered as the X-ray binaries in the `very high state', while mechanisms are relatively unclear due to the short duration. The~steady jets in quiescent and low/hard state are generally believed to be launched by extracting the energy of either an accretion disk (the `BP' jets,~\citep{1982MNRAS.199..883B}) or the spin of the central compact object (the `BZ' jets,~\citep{1977MNRAS.179..433B}), while episodic jets are alternatively explained by other models (e.g., the instability of accretion disks~\citep{2019ApJ...877..130S}).

Several works indicate that the accretion in the transition state can be described as a `slim disk'~\citep{2013MNRAS.436...71V}, which supports the near and moderately super-Eddington accretion. As~the super-Eddington accretion has an essential role in ultraluminous X-ray sources and high redshift quasars,  the transition state is also named an `ultraluminous state'~\cite{2009MNRAS.397.1836G}. Due to the bright episodic jet eruption (which is brighter than steady jets), the~radio emissions observed in IMBHs and low-luminosity AGNs likely correspond to the ejecta of the `ultraluminous state'. However, the~understanding of the accretion and ejection properties in the `ultraluminous state' is limited to only a few XRBs, e.g.,~Cygnus X-3, GRS 1915 + 105 and~SS\,433.

It was theoretically studied that the accretion flow in supermassive black holes in active galactic nuclei resembles the stellar mass black holes in X-ray binary systems. Therefore, the two systems, even though they have a mass range of 10 dexes, should share the same correlations. Indeed, there are several correlations that work in both stellar mass and supermassive black holes (e.g.,~\citep{2003MNRAS.345.1057M, 2017A&A...603A.127S, 2020ApJ...904..200Y}). The~universal correlations among the diverse accretion systems indicate a similar physics in accretion, that naturally induces the scale-invariant accretion and ejection phenomena in both radio and X-ray. Historically, there are several correlations that are successful in unifying XRBs and AGNs: (1) the most applicable correlation among X-ray luminosity, radio luminosity and~black hole mass was named  the fundamental plane of black hole activity~\citep{2003MNRAS.345.1057M, 2004A&A...414..895F}. The~fundamental plane correlation was initially identified in X-ray binaries when their accretion state is in the low/hard state, where the accretion is in low and moderate rates. As~the accretion transits to the `high/soft' state, the~radio emission (thus the jet) dramatically drops while the X-ray emission remains in a high luminosity, which deviates from the fundamental plane relation. The~similarity between XRBs and AGNs not only indicated that both of them satisfy the fundamental plane relation of black hole activity but also the similar radio quenching as the increase of accretion rates. At~present, the~fundamental plane relation has a clear gap between the stellar mass and supermassive black holes. Filling the gap would build a more fundamental and comprehensive understanding of the accretion systems; therefore, IMBHs are the key to linking the stellar mass and supermassive black holes in the fundamental plane relation. The~fundamental plane relation is important as it links both accretion and ejection processes and~is applicable based on X-ray and radio surveys; (2) because the fundamental plane relation of black hole activity only works well in the low/hard state, a universal correlation developed, i.e.,~the inverse correlation between the radio loudness and the Eddington ratio~\citep{2002ApJ...564..120H, 2007ApJ...658..815S, 2011MNRAS.417..184B, 2020ApJ...904..200Y}. The~inverse correlation indicates the suppression of the jet producing as the increase of Eddington ratios, which is applicable for the radio emission produced in the high and super-Eddington XRBs and AGNs; (3)~the correlation among the bolometric luminosity, black hole mass and~the characteristic timescales of X-ray variability~\cite{2006Natur.444..730M}; (4) the hardness--intensity diagram (HID,~\citep{2006MNRAS.372.1366K}) and the accretion rate ratio--intensity diagram (ARRID,~\citep{2014ApJ...786....4M, 2016ApJ...819..107J}) for individual XRB and~the statistical version, the~ disk-fraction luminosity diagrams~\cite{2006MNRAS.372.1366K, 2017A&A...603A.127S}; (5) the inverse correlation between QPO frequencies and black hole masses~\citep{2004ApJ...609L..63A, 2015ApJ...798L...5Z, 2015ApJ...811L..11P}. However, evidence is still lacking that the above correlations are applicable for all accreting systems, i.e.,~these correlations need to be explored in the mass scale of $10^3$--$10^5\,M_\odot$.

\section{Radio Activities of~IMBHs}
Given that dwarf galaxies tend to host low-mass AGNs (or IMBHs), \citet{2020ApJ...888...36R} conducted a radio census of  111 FIRST detected dwarf galaxies (the stellar mass range from $3\times10^7$ to $3\times10^9\,M_\odot$). Among~them, only 13 were identified to be powered by AGNs, through radio diagnoses, e.g.,~the compact point-like morphologies and the excess radio emissions compared with those afforded by star-forming activities. This yields an AGN rate of 11\% in radio-active dwarf galaxies. Only one of these 13 was cross identified in optical bands and~therefore it results in 0.9\%  AGN fraction in dwarf galaxies simultaneously. However, the~fraction may not imply a physical distribution due to the sample selection effect and the incompleteness of the~sample.

Alternatively, we~\citep{yang2022radio} compiled a sample of 598 IMBH candidates which were uniformly selected from the Sloan Digital Sky Survey (SDSS,~\citep{2004ApJ...610..722G,2007ApJ...670...92G, 2012ApJ...755..167D, 2013ApJ...775..116R, 2018ApJS..235...40L}). Through cross-matching with the NRAO VLA Sky Survey (NVSS,~\citep{1998AJ....115.1693C}) and Faint Images of the Radio Sky at Twenty centimeters (FIRST,~\citep{1995ApJ...450..559B}) catalogs within 1 arcsec of their optical positions, \citet{yang2022radio} found that 36 sources (6\%) have radio counterparts (with signal-to-noise ratios~$>$9) in the FIRST survey. Furthermore, we selected four IMBH candidates (0.6\%) for VLBI observations with the further criteria: (1) the estimated black hole mass should be $<$$10^6 M_\odot$, (2) the 1.4\,GHz radio flux density of the FIRST survey is $>$2 mJy, which yields a signal-to-noise ratio of $>$12, and~(3) objects associated with the point X-ray emissions. \citet{yang2022radio} successfully detected parsec-scale radio emissions in these four candidates and confirmed that the IMBH candidates are radio-active, i.e.,~they maintain ejection processes. In~Table~\ref{tab1}, we summarize the IMBH candidates with both radio and X-ray~observations.

Comparing the optical and radio emission regions of these IMBH candidates, most of the radio-active IMBHs have radio emission regions offset from the optical nuclei. This indicates that the radio emissions in these cases are more likely relics of ejecta than radio-active nuclei. Given the small fraction of radio-active IMBHs and the lesser rate of actively ejecting ones of them, it satisfies the scenario of episodic ejections and a long evolution cycle from a quiescent state to a very high state. The~episodic ejection nature in IMBHs has been further identified in a few other candidates with multiple radio-emitting blobs, e.g.,~NGC~4293~\citep{2022MNRAS.517.4959Y}, NGC~4395~\citep{2022MNRAS.514.6215Y} and RGG9~\citep{2020MNRAS.495L..71Y}.

In X-ray binaries, an~episodic ejection is produced when accretion is in a very high/inter\\-mediate state~\citep{2004MNRAS.355.1105F}. Interestingly, an~episodic ejection was already observed in the IMBH candidate HLX-1 when it was in a high X-ray state~\citep{2012Sci...337..554W}. Based on the unification model of accretion~\citep{2006Natur.444..730M}, AGNs should also have specific accretion states and relevant state transitions as what is universally observed in Galactic XRBs~\citep{2004MNRAS.355.1105F}. Naturally, accretion states should also be expected in IMBHs and~they should evolve much faster than  in AGNs (as the timescale is proportional to black hole masses, e.g.,~\citep{2017A&A...603A.127S}). 

\begin{table}[H] 
\caption{IMBH candidates with both radio and X-ray~observations. \label{tab1}}
\newcolumntype{C}{>{\centering\arraybackslash}X}
\begin{tabularx}{\textwidth}{cCCCC}
\toprule
\textbf{Name}	  & \boldmath{$\log{M}$}    & \boldmath{$\log{L_\mathrm{R,VLBI}}$}  & \boldmath{$\log{L_\mathrm{R,VLA}}$}     & \boldmath{$\log{L_X}$}        \\
 \textbf{(SDSS or Alias)}       & \textbf{(}\boldmath{$M_\odot$}\textbf{)}  & \textbf{(erg\,s}\boldmath{$^{-1}$}\textbf{)}   & \textbf{(erg\,s}\boldmath{$^{-1}$}\textbf{)}       & \textbf{(erg\,s}\boldmath{$^{-1}$}\textbf{)}   \\
\midrule
J024656.39 $-$ 003304.8  &$5.7~^\dagger $&$                         $&$ <$$36.34~^\alpha $&$39.8\pm0.6~^a$      \\
J090613.76 $+$ 561015.1  &$5.6~^\dagger $&$   38.04\pm0.03~^\delta   $&$ 38.66\pm0.02~^\alpha $&$40.1\pm0.4~^a$      \\
J095418.15 $+$ 471725.1  &$4.9~^\dagger $&$                         $&$ 36.73\pm0.06~^\alpha $&$40.1\pm0.4~^a$      \\
J144012.70 $+$ 024743.5  &$5.2~^\dagger $&$                         $&$ 37.83\pm0.01~^\alpha $&$39.8\pm0.6~^a$      \\
J152637.36 $+$ 065941.6  &$5.5~^\dagger $&$                         $&$ 36.68\pm0.10~^\alpha $&$41.84\pm0.04~^a$    \\
J160531.84 $+$ 174826.1  &$5.2~^\dagger $&$                         $&$ <$$35.99~^\alpha $&$41.29\pm0.08~^a$    \\
J082443.28 $+$ 295923.5  &$5.6~^\ddagger$&$   36.77 \pm 0.08~^\beta  $&$ 38.00 \pm 0.03~^\beta $&$ 42.4~^b$ \\
J110501.98 $+$ 594103.5  &$5.5~^\ddagger$&$   37.84 \pm 0.03~^\beta  $&$ 38.10 \pm 0.04~^\beta $&$ 42.1~^b$ \\
J131659.37 $+$ 035319.9  &$5.8~^\ddagger$&$   37.80 \pm 0.05~^\beta  $&$ 38.26 \pm 0.08~^\beta $&$ 41.7~^b$ \\
J132428.24 $+$ 044629.6  &$5.7~^\ddagger$&$   37.09 \pm 0.06~^\beta  $&$ 37.46 \pm 0.06~^\beta $&$ 41.7~^b$ \\
J122112.82 $+$ 182257.7  &$5.53~^\bullet$&$  <$$35.46~^\gamma          $&$  36.18 \pm 0.04~^\gamma $&$  39.69~^c $ \\
J111552.01 $-$ 000436.1  &$5.05~^\bullet$&$  <$$36.52~^\epsilon          $&$  38.73 \pm 0.03~^\epsilon        $&$  40.6~^d $ \\
NGC~404                &$5.74~^\star$&$  <$$33.30~^\theta           $&$  34.58~^\zeta         $&$  37.14~^e $ \\
NGC~4395               &$3.95~^\circ$&$  33.67~^\iota           $&$  34.10~^\eta         $&$  40.0~^f $ \\
\bottomrule
\end{tabularx}
\noindent{\footnotesize{$\dagger$: \citet{2013ApJ...775..116R}; $\ddagger$: \citet{2007ApJ...670...92G}; $\bullet$: \citet{2018ApJ...863....1C}; $\star$: \citet{2020MNRAS.496.4061D}; $\circ$: \citet{2019NatAs...3..755W}; $a$:~\citet{2017ApJ...836...20B}; $b$: \citet{2014ApJ...788L..22G}; $c$: \citet{2022MNRAS.512.3284S}; $d$: \citet{2018ApJ...863....1C}; $e$: \citet{2014ApJ...791....2P}; $f$:~\citet{2005AJ....129.2108M}; $\alpha$: \citet{2022MNRAS.516.6123G}; $\beta$: \citet{yang2022radio}; $\gamma$: \citet{2022MNRAS.517.4959Y}; $\delta$: \citet{2020MNRAS.495L..71Y}; $\epsilon$: this work; $\zeta$:~\citet{2012ApJ...753..103N}; $\eta$:~\citet{2018A&A...616A.152S}; $\theta$: \citet{2014ApJ...791....2P}; $\iota$: \citet{2022MNRAS.514.6215Y}. Note that a few sources have only VLA observations, i.e.,~VLBI luminosity is not available.}}

\end{table}

\section{The Black Hole Fundamental Plane of Involving~IMBHs}
Accretion onto compact objects follows the so-called fundamental plane relation (FMP) of black hole activity, which is among nuclear radio and X-ray luminosities and~black hole masses of any accretion systems. The~fundamental plane relation originates from jet--disk coupling, where the jet produces radio emission and the accretion disk and corona system produce X-ray emission and~the mass of the central engine regulates both the accretion and ejection power. The~correlation between radio and X-ray luminosity was initially inferred from X-ray binary systems that host stellar mass black holes, which only work in quiescent and low/hard accretion states that are associated with steady ejections~\citep{2003MNRAS.344...60G, 2003MNRAS.345.1057M}. 

In addition, the~very high/intermediate state will substantially trigger episodic radio ejection, then subsequently the radio emission will dramatically decrease as the accretion state evolves from the very high state to the soft state. Similarly, we expect that the radio power of blobs may also be correlated with the X-ray luminosity. Indeed, several works (e.g.,~\citep{2003MNRAS.345.1057M, 2003MNRAS.344...60G}) explored the radio-to-X-ray correlation in transient (very high) states, in~X-ray binaries; the results indicate that the correlation may still be maintained but with high dispersion. It was already known that the fundamental plane of black hole activity can be contaminated by several mechanisms, e.g.,~the enhancement of the radio emission of lobes (i.e., the shock waves) when propagating through a dense medium~\citep{2021A&ARv..29....3O, 2022MNRAS.517.4959Y}; the radio contamination by AGN hosts; the relativistic beaming effect of jets; the X-ray contamination by the jet~\citep{2008ApJ...680..911S}. Therefore, the~fundamental plane relation works in both low/hard and very high/intermediate states. The~fundamental plane of black hole activity presented by \citet{2003MNRAS.345.1057M} is
\begin{equation}
\log{L_R}=(0.60^{+0.11}_{-0.11})\log{L_X}+(0.78^{+0.11}_{-0.09})\log{M_\mathrm{BH}}+7.33^{+4.05}_{-4.07}. 
\end{equation}
There are also recent updates, e.g., \citet{2018A&A...616A.152S}
\begin{equation}
\log{L_R}=(0.48\pm0.04)\log{L_X}+(0.79\pm0.03)\log{M_\mathrm{BH}}+11.71, 
\end{equation}
and \citet{2019ApJ...871...80G}

\begin{adjustwidth}{-\extralength}{0cm}
\begin{equation}
\begin{aligned}
& \log{\left(\frac{M_\mathrm{BH}}{10^8\,M_\odot}\right)}= (1.09\pm0.10)\log{\left(\frac{L_R}{10^{38}\,\mathrm{erg\,s^{-1}}}\right)}+\left(-0.59^{+0.16}_{-0.15}\right)\log{\left(\frac{L_X}{10^{40}\,\mathrm{erg\,s^{-1}}}\right)}+(0.55\pm0.22).
\end{aligned}
\end{equation}
\end{adjustwidth}
Here we only explore the fundamental plane from \citet{2003MNRAS.345.1057M} based on the scope of this work, i.e.,~the radio emission of AGNs (especially low luminosity and low-mass AGNs) should be obtained from the parsec-scale region and the comparison sample used in this work is from the same work~\citep{2003MNRAS.345.1057M}.

\begin{figure}[H]
\includegraphics[width=8.5 cm]{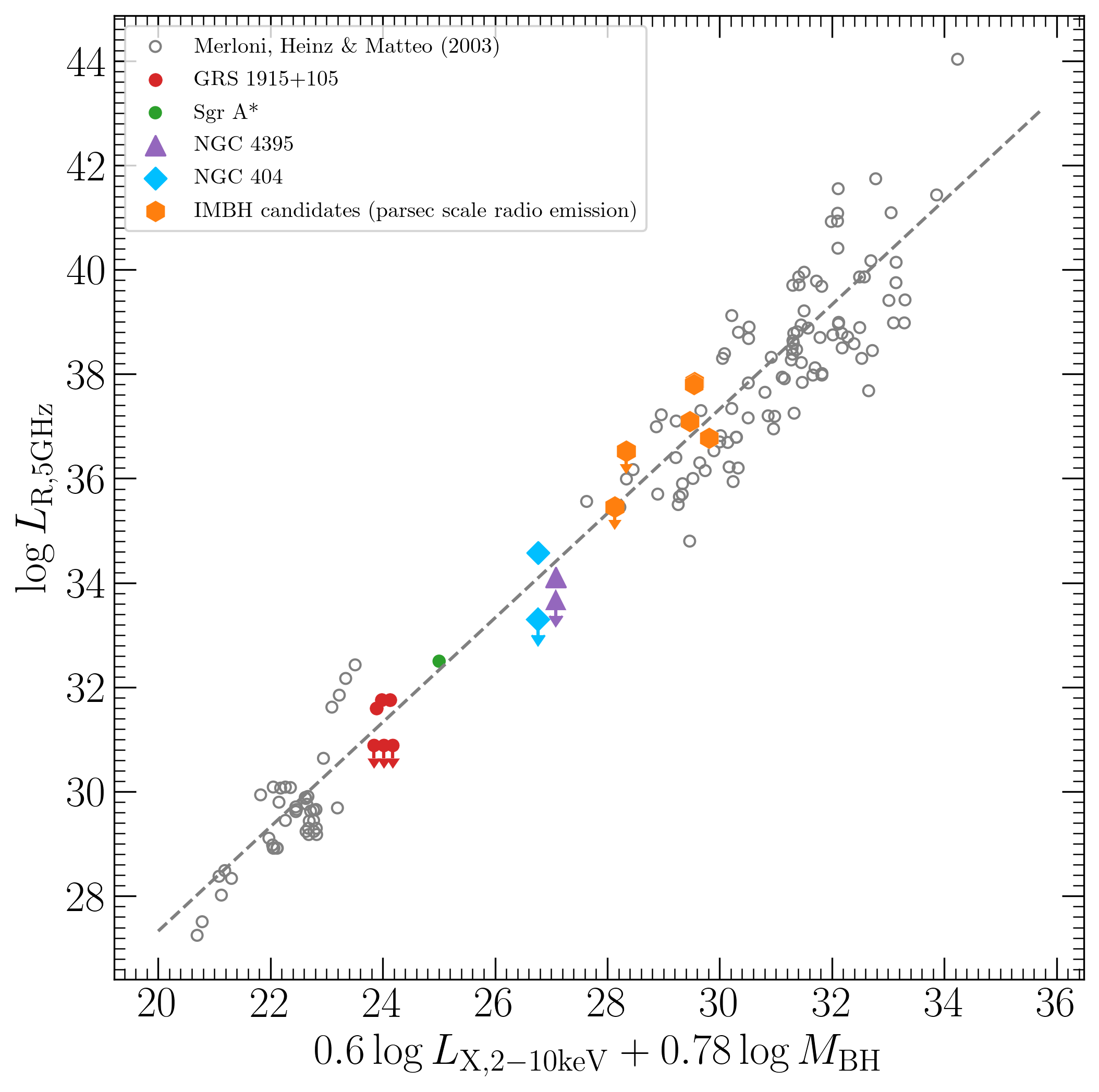}
\caption{The fundamental plane relation of black hole activity with the parsec scale radio luminosity of IMBH candidates. The~XRBs, data for Sgr A$^\ast$ and GRS 1915 + 105 and~AGNs are from \citet{2003MNRAS.345.1057M} and~the IMBH candidates are from \citet{yang2022radio} and \citet{2022MNRAS.517.4959Y}.\label{fig1}}
\end{figure}

\begin{figure}[H]
\includegraphics[width=8.5 cm]{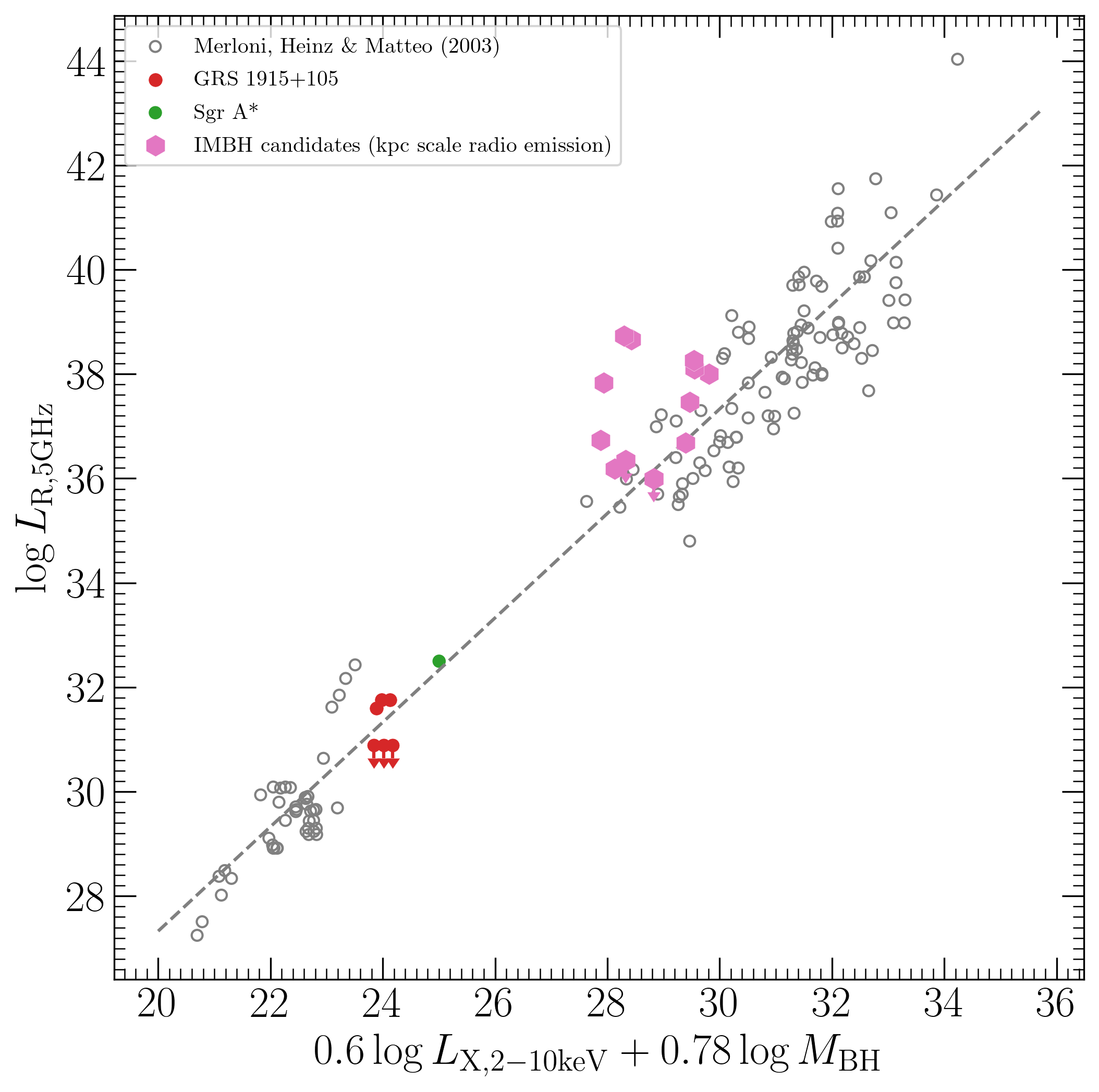}
\caption{The fundamental plane relation of black hole activity with the kpc scale radio luminosity of IMBH candidates. The~radio luminosity from the kpc scale is clearly overestimated with regard to the fundamental~plane. \label{fig2}}
\end{figure}

The fundamental plane of black hole activity indicates a scale-free accretion and ejection physics in both SBHs and AGNs. Recently, \citet{yang2022radio} explored the fundamental plane relation in a sample of IMBH candidates with pc-scale radio emissions obtained from VLBI observations; which implies that IMBHs follow the fundamental plane of black hole activity. In~Figure~\ref{fig1}, we use the IMBH sample from \citet{yang2022radio} and \citet{2022MNRAS.517.4959Y}. There is one more IMBH candidate (J111552.01 $-$ 000436.1) from the same observation of \citet{yang2022radio} but it was not detected; here, we take its upper limit. Furthermore, we involve a few IMBH candidates from the literature (see comments in Table~\ref{tab1}). In particular, we only use the radio emission from VLBI observations (see Table~\ref{tab1}) to reduce the contamination from hosts. However, it is hard to determine whether to take the radio emissions from lobes or core regions. Neither the radio emission from lobes nor core regions indicate the power of ejection by considering that the lobe may receive enhancement by circumnuclear medium and beaming effect and~the effect of episodic ejection and state transition. To~reduce the deficit (see Figure \ref{fig2}), we use the radio emissions close to the core region, which are accompanied by the core X-ray emissions. Specifically, we use the radio emission from the core region of NGC\,4293 as the lobes of this source received a strong enhancement by circumnuclear medium (see~\citep{2022MNRAS.517.4959Y}). The~linear distribution trend presented in Figure~\ref{fig1} indicates that the accretion and ejection processes in IMBHs are similar to those in SBHs and SMBHs, i.e.,~there exists a unified model for the accretion and ejection process for SBHs, IMBHs and~SMBHs. Obviously, involving IMBHs in the fundamental plane fills the gap between the stellar mass and supermassive black holes, i.e.,~low luminosity AGNs (powered by supermassive black holes) extend to the lower-left in the fundamental plane relation and approximately overlap with the highly accreting IMBHs and~low-rate accreting IMBHs touch the most luminous stellar-mass black~holes.

\unskip

We will emphasize that the fundamental plane relation is devoted to depicting the correlation between the radio and X-ray luminosity in the ejection phase (excepting the canonical `soft state'), which is relatively difficult for IMBHs and low-mass AGNs. Radio observations of these sources are generally limited to the radio emissions produced from very high states because the radio emissions in quiescent and low/hard states are likely dominated by the compact and steady jet, which is weaker than the episodic radio blobs in very high states. However, we do not track the evolution of each radio blob, but~the radio luminosity of the instance ejection and associated X-ray luminosity. Again, low-resolution radio observations may collect the radio emissions from multiple blobs of one or more evolution cycles, which will obviously overestimate the radio emission from one single blob. Due to the long timescale of the evolution cycle, it is generally impossible to record the radio and X-ray emission in the very high state of individual IMBH. Fortunately, the~radio blobs in IMBHs and low-mass AGNs will live for a sufficiently longer time than in XRBs and move to a distance from parsec to kilo-parsec scales. Therefore, it is applicable to take the radio emission from a recent blob along with the core X-ray~emission.

In Figure~\ref{fig1}, we also involve pc and sub-pc-scale radio constraints of two IMBH candidates, NGC~4395 and NGC~404, having low accretion rates. They have the Eddington ratios of $\lambda_\mathrm{Edd}=1.5\times10^{-6}$~\citep{2014ApJ...791....2P} and $1.2\times10^{-3}$~\citep{2006ApJ...646L..95W}, respectively. In~this plot, the~radio luminosity of NGC~404 was taken from the nuclear 7\,parsec scale region through VLA A-array 5\,GHz observation~\citep{2012ApJ...753..103N} and~we estimated the 5\,GHz radio luminosity of NGC~4395 via VLA A-array 15\,GHz observation~\citep{2018A&A...616A.152S}, from~the nuclear 4\,parsec scale region. Both the radio luminosities of the above two sources are obtained from the parsec-scale region, which is comparable to the VLBI observations of the other IMBHs presented in this plot due to the low redshift of NGC~4395 and NGC~404. Furthermore, both NGC~4395 and NGC~404 received the new measurements of black hole masses; see \citet{2019NatAs...3..755W} and~\mbox{\citet{2020MNRAS.496.4061D}}, respectively. In particular, we take the X-ray luminosity of NGC~404 from $Chandra$ observation~\citep{2014ApJ...791....2P}. It should be noted here that the~two sources have radio emission at parsec scales, but~VLBI observations only reveal the upper limit of them. Interestingly, a~sample of low luminosity AGNs has a similar radio drop from parsec to sub-parsec scales~\cite{2021ApJ...906...88F}. Certainly, the~non-detection of sub-parsec scale radio emissions in a few low accretion rate AGNs~\cite{2021ApJ...906...88F} and IMBHs (NGC~4395 and NGC~404) challenges the (Merloni) fundamental plane and~deeper VLBI observations are required to unveil the nature of these central engines. Taking the unified scheme for the accretion and ejection process, it was already shown that the episodic radio ejection and the high Eddington ratios in a few \mbox{IMBHs~\citep{2020ApJ...904..200Y, 2020MNRAS.495L..71Y, 2022MNRAS.514.6215Y, yang2022radio, 2022MNRAS.517.4959Y}} draw a similarity with the canonical very high/intermediate state in XRBs. On~the other hand, the~low Eddington ratios in a few low luminosity AGNs~\cite{2021ApJ...906...88F} and IMBHs (NGC~4395 and NGC~404) alternatively indicate quiescent or low/hard accretion states. Studying XRBs shows that radio emissions will drop with the transition from a very high/intermediate state to a soft state and~ comparing between a very high state and a quiescent state~\citep{2000A&A...359..251C}. In~Figure~\ref{fig1}, we mark the monitoring observations of the microquasar GRS~1915 + 105 to illustrate the similarity and it is likely consistent with the observational signatures in NGC~4395 and NGC~404.

\section{The Correlation between the Size Scale of Outflows and the Masses of~Accretors}
Several critical factors regarding the size scales of outflows include the power of the central engine and the external environment. Given that external environments vary in each object, the~largest scale of outflows should be only related to the power or type of the central engines/accretors for ejecting objects with similar ages, statistically. Again, the~distribution of ejecting ages directly results in the positive distribution of outflow sizes and~the largest outflow size tends to be associated with the older (over a critical age) ones of them. Naturally, the~largest size scale of outflow-dominated feedback relies on the power of the central engine, hence the mass of the central engine according to the unified accretion and ejection process. Therefore, a~positive correlation between the largest (linear size) scale of outflows $\hat{LS}_\mathrm{out}$ and the masses of accretors $M_\mathrm{core}$ should be~expected.

\begin{figure}[H]
\includegraphics[width=8.5 cm]{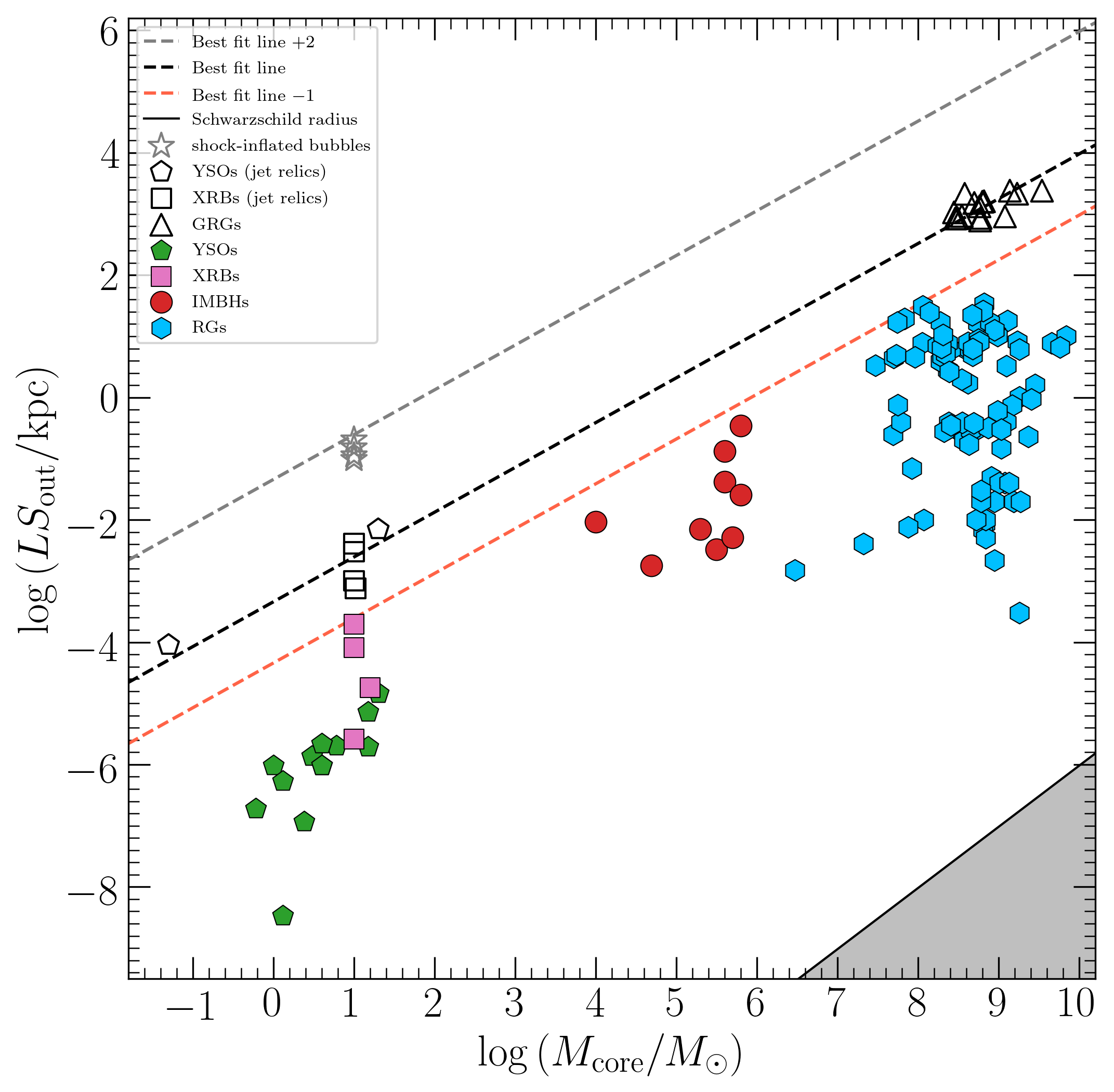}
\caption{The linear size of outflow distribution along the masses of~accretors. \label{fig3}}
\end{figure}

Now we try to explore the $M_\mathrm{core}-LS_\mathrm{out}$ correlation among a sample of accretors with masses ranging from $10^{-2}$ to $10^{10}\,M_\odot$ (the data will be presented in a forthcoming paper, \citet{yang2023}, along with additional samples). This sample includes young stellar objects (YSOs), X-ray binary systems (XRBs), intermediate-mass black holes (IMBHs), normal radio galaxies (RGs) and~giant radio galaxies (GRGs). Obviously, the $\hat{LS}_\mathrm{out}-M_\mathrm{core}$ correlation can be identified by inspecting the upper envelope of jet sizes (the line `best fit $-1$' in Figure~\ref{fig3}). However, jets can further extend to form relics and bubbles; on the largest scale of outflows, the~jet relics in GRGs and a few YSOs and XRBs may represent the largest structures of the ejected plasma (fossil plasma); in addition, the~shocks driven by outflows tend to form a more elongated structure (i.e., shock-inflated bubbles) than fossil plasma (see~\citep{yang2023}). Interestingly, the~jet relics in GRGs and a few YSOs and XRBs follow the trend identified in (active) jets, which ensures we fit a linear correlation between the size of jet relics and the masses of accretors and gives the following parameters,
\begin{equation}
\log{\hat{LS}_\mathrm{out}} = (0.73\pm0.01)\log{M_\mathrm{core}} - (3.34\pm0.10), 
\end{equation}
where $\hat{LS}_\mathrm{out}$ is in kpc and $M_\mathrm{core}$ is in solar~mass.

The distribution between the linear size of outflows and the masses of accretors is presented in Figure~\ref{fig3}. We further extend the linear correlation by $-$1 and 2 dex, which likely links the largest size scale of active jets and shock-inflated bubbles, respectively. This implies that the shock-inflated bubbles can reach a 3-dex longer extension than the active jets. Accordingly, a~hypothesis is that the jets in IMBH may provide a few hundred kilo-parsec scale feedback, which is beyond the size of their dwarf hosts. Again, based on the size of the shock-inflated bubbles in XRBs, jets produced by XRBs will drive shocks to further blow bubbles of 2-dex larger than the jet relics. If~this scaling relation holds for both IMBHs and SMBHs, then we expect to find shock-inflated bubbles of 10--100\,kpc and~100\,Mpc scale in IMBHs and the most massive SMBHs, respectively. Furthermore, $LS_\mathrm{out}-M_\mathrm{core}$ correlation lacks the data of jet relics in the IMBH mass scale, which requires further exploration in observation. Again, a~theoretical deduction is required as well (see a forthcoming paper, \citet{yang2023}).

The kinetic power is the key parameter to drive jet expansion, which is thought to be correlated with the black hole spin and mass~\citep{1977MNRAS.179..433B, 1982MNRAS.199..883B, 2007MNRAS.377.1652N} and the structure of the accretion disk~\citep{1995ApJ...452..710N, 2009MNRAS.395.2183Y, 2019ApJ...877..130S}. The~correlation between the jet kinetic power and black hole masses is indicated in several statistical analyses (e.g.,~\citep{2012Sci...338.1445N, 2021ApJ...908...85M, 2023MNRAS.519.6199C}), while the observational correlation between the jet power and black hole spin is still a mystery (e.g.,~\citep{2021ApJ...913...93C}). Except for the kinetic power, the~size scale of an outflow in an individual object may be regulated by ages~\citep{2021A&ARv..29....3O}, inclination angle and interplay between jets and surrounding media~\citep{2018MNRAS.477.2128H}. These parameters insert dispersion in the $LS_\mathrm{out}$ distribution along the mass of accretors (see, e.g.,~\citep{2007MNRAS.377.1652N}). However, the~largest size scale of outflow (the upper envelope in $LS_\mathrm{out}-M_\mathrm{core}$ distribution) is likely free from those parameters due to the selection effect, i.e.,~the largest scale of outflow tends to be associated with the longest-lasting and the highest spin central engines. Again, a~low frustration from the surrounding media is an assistance to allow the jet expansion further. Therefore, the correlation between the size scale of outflows and the masses of~accretors can be considered to be intrinsic.

\vspace{6pt} 

\authorcontributions{Conceptualization, X.Y. and J.Y.; methodology, X.Y.; software, X.Y.; validation, X.Y. and J.Y.; formal analysis, X.Y.; investigation, X.Y.; resources, X.Y.; data curation, X.Y.; writing---original draft preparation, X.Y.; writing---review and editing, X.Y.; visualization, X.Y.; supervision, X.Y.; project administration, X.Y.; funding acquisition, X.Y. All authors have read and agreed to the published version of the manuscript. }

\funding{This research was funded by the Shanghai Sailing Program grant number 21YF1455300, China Postdoctoral Science Foundation grant number 2021M693267 and~the National Science Foundation of China grant number 12103076.}

\dataavailability{This work has made use of data from the VLBA project BA146 (available in the NRAO Data Archive: \url{https://data.nrao.edu/}) and the EVN project EY039 (available in the EVN Data Archive at JIVE: \url{http://archive.jive.nl/scripts/portal.php}).}

\acknowledgments{The National Radio Astronomy Observatory is a facility of the National Science Foundation operated under cooperative agreement by Associated Universities, Inc. The EVN is a joint facility of independent European, African, Asian, and North American radio astronomy institutes.}

\conflictsofinterest{The authors declare no conflict of interest.} 

\newpage
\abbreviations{Abbreviations}{
The following abbreviations are used in this manuscript:\\

\noindent 
\begin{tabular}{@{}ll}
AGNs   &   Active Galactic Nuclei \\
IMBHs  &   Intermediate-mass black holes \\
SBHs   &   Stellar-mass black holes \\
SMBHs  &   Supermassive black holes \\
XRBs   &   X-ray binaries \\
YSOs   &   Young Stellar Objects \\
GRGs   &   Giant Radio Galaxies \\
RQ-AGNs&   Radio-quiet AGNs \\
RL-AGNs&   Radio-loud AGNs \\
VLBI   &   Very Long Baseline Interferometry \\
VLBA   &   the Very Long Baseline Array \\
VLA    &   the Very Large Array \\
EVN    &   the European VLBI Network
\end{tabular}
}

\begin{adjustwidth}{-\extralength}{0cm}
\printendnotes[custom]

\reftitle{References}

\end{adjustwidth}

\begin{thebibliography}{999}

\bibitem[{Mirabel}(2017)]{2017NewAR..78....1M}
{Mirabel}, F.
\newblock {The formation of stellar black holes}.
\newblock {\em New Astron. Rev.} {\bf 2017}, {\em 78},~1--15.

\bibitem[{Kormendy} and {Ho}(2013)]{2013ARA&A..51..511K}
{Kormendy}, J.; {Ho}, L.C.
\newblock {Coevolution (Or Not) of Supermassive Black Holes and Host Galaxies}.
\newblock {\em Annu. Rev. Astron. Astrophys.} {\bf 2013}, {\em 51},~511--653.

\bibitem[{Greene} et~al.(2020){Greene}, {Strader} and
  {Ho}]{2020ARA&A..58..257G} Greene, J.E.; Strader, J.; Ho, L.C.
\newblock {Intermediate-Mass Black Holes}.
\newblock {\em Annu. Rev. Astron. Astrophys.} {\bf 2020}, {\em 58},~257--312. \newblock {\url{https://doi.org/10.1146/annurev-astro-032620-021835}}.

\bibitem[{Wu} et~al.(2015){Wu}, {Wang}, {Fan}, {Yi}, {Zuo}, {Bian}, {Jiang},
  {McGreer}, {Wang}, {Yang}, {Yang}, {Thompson} and
  {Beletsky}]{2015Natur.518..512W}
{Wu}, X.B.; {Wang}, F.; {Fan}, X.; {Yi}, W.; {Zuo}, W.; {Bian}, F.; {Jiang},
  L.; {McGreer}, I.D.; {Wang}, R.; {Yang}, J.;  et~al.
\newblock {An ultraluminous quasar with a twelve-billion-solar-mass black hole
  at redshift 6.30}.
\newblock {\em Nature} {\bf 2015}, {\em 518},~512--515.

\bibitem[{Ba{\~n}ados} et~al.(2018){Ba{\~n}ados}, {Venemans}, {Mazzucchelli},
  {Farina}, {Walter}, {Wang}, {Decarli}, {Stern}, {Fan}, {Davies}, {Hennawi},
  {Simcoe}, {Turner}, {Rix}, {Yang}, {Kelson}, {Rudie} and
  {Winters}]{2018Natur.553..473B}
{Ba{\~n}ados}, E.; {Venemans}, B.P.; {Mazzucchelli}, C.; {Farina}, E.P.;
  {Walter}, F.; {Wang}, F.; {Decarli}, R.; {Stern}, D.; {Fan}, X.; {Davies},
  F.B.;  et~al.
\newblock {An 800-million-solar-mass black hole in a significantly neutral
  Universe at a redshift of 7.5}.
\newblock {\em Nature} {\bf 2018}, {\em 553},~473--476.

\bibitem[{Volonteri}(2010)]{2010A&ARv..18..279V}
{Volonteri}, M.
\newblock {Formation of supermassive black holes}.
\newblock {\em  Astron. Astrophys. Rev.} {\bf 2010}, {\em 18},~279--315.

\bibitem[{Inayoshi} et~al.(2020){Inayoshi}, {Visbal} and
  {Haiman}]{2020ARA&A..58...27I}
{Inayoshi}, K.; {Visbal}, E.; {Haiman}, Z.
\newblock {The Assembly of the First Massive Black Holes}.
\newblock {\em Annu. Rev. Astron. Astrophys.} {\bf 2020}, {\em 58},~27--97.
\newblock {\url{https://doi.org/10.1146/annurev-astro-120419-014455}}.

\bibitem[{Mezcua}(2017)]{2017IJMPD..2630021M}
{Mezcua}, M.
\newblock {Observational evidence for intermediate-mass black holes}.
\newblock {\em Int. J. Mod. Phys. D} {\bf 2017}, {\em
  26},~1730021.

\bibitem[{Graham} and {Soria}(2019)]{2019MNRAS.484..794G}
{Graham}, A.W.; {Soria}, R.
\newblock {Expected intermediate-mass black holes in the Virgo cluster---I.
  Early-type galaxies}.
\newblock {\em Mon. Not. R. Astron. Soc.} {\bf 2019}, {\em 484},~794--813.
\newblock {\url{https://doi.org/10.1093/mnras/sty3398}}.

\bibitem[{Graham} et~al.(2019){Graham}, {Soria} and
  {Davis}]{2019MNRAS.484..814G}
{Graham}, A.W.; {Soria}, R.; {Davis}, B.L.
\newblock {Expected intermediate-mass black holes in the Virgo cluster---II.
  Late-type galaxies}.
\newblock {\em Mon. Not. R. Astron. Soc.} {\bf 2019}, {\em 484},~814--831.
\newblock {\url{https://doi.org/10.1093/mnras/sty3068}}.

\bibitem[{Secrest} et~al.(2012){Secrest}, {Satyapal}, {Gliozzi}, {Cheung},
  {Seth} and {B{\"o}ker}]{2012ApJ...753...38S}
{Secrest}, N.J.; {Satyapal}, S.; {Gliozzi}, M.; {Cheung}, C.C.; {Seth}, A.C.;
  {B{\"o}ker}, T.
\newblock {The Chandra View of NGC 4178: The Lowest Mass Black Hole in a
  Bulgeless Disk Galaxy?}
\newblock {\em  Astrophys. J.} {\bf 2012}, {\em 753},~38.
\newblock {\url{https://doi.org/10.1088/0004-637X/753/1/38}}.

\bibitem[{Mezcua} et~al.(2015){Mezcua}, {Roberts}, {Lobanov} and
  {Sutton}]{2015MNRAS.448.1893M}
{Mezcua}, M.; {Roberts}, T.P.; {Lobanov}, A.P.; {Sutton}, A.D.
\newblock {The powerful jet of an off-nuclear intermediate-mass black hole in
  the spiral galaxy NGC 2276}.
\newblock {\em Mon. Not. R. Astron. Soc.} {\bf 2015}, {\em 448},~1893--1899.
\newblock {\url{https://doi.org/10.1093/mnras/stv143}}.

\bibitem[{Yang} et~al.(2022){Yang}, {Wang} and {Guo}]{2022MNRAS.517.4959Y}
{Yang}, X.; {Wang}, R.; {Guo}, Q.
\newblock {A compact symmetric ejection from the low mass AGN in the LINER
  galaxy NGC 4293}.
\newblock {\em Mon. Not. R. Astron. Soc.} {\bf 2022}, {\em 517},~4959--4967.
\newblock {\url{https://doi.org/10.1093/mnras/stac2990}}.

\bibitem[{Kaspi} et~al.(2000){Kaspi}, {Smith}, {Netzer}, {Maoz}, {Jannuzi} and
  {Giveon}]{2000ApJ...533..631K}
{Kaspi}, S.; {Smith}, P.S.; {Netzer}, H.; {Maoz}, D.; {Jannuzi}, B.T.;
  {Giveon}, U.
\newblock {Reverberation Measurements for 17 Quasars and the
  Size-Mass-Luminosity Relations in Active Galactic Nuclei}.
\newblock {\em  Astrophys. J.} {\bf 2000}, {\em 533},~631--649.
\newblock {\url{https://doi.org/10.1086/308704}}.

\bibitem[{Woo} et~al.(2019){Woo}, {Cho}, {Gallo}, {Hodges-Kluck}, {Le}, {Shin},
  {Son} and {Horst}]{2019NatAs...3..755W}
{Woo}, J.H.; {Cho}, H.; {Gallo}, E.; {Hodges-Kluck}, E.; {Le}, H.A.N.; {Shin},
  J.; {Son}, D.; {Horst}, J.C.
\newblock {A 10,000-solar-mass black hole in the nucleus of a bulgeless dwarf
  galaxy}.
\newblock {\em Nat. Astron.} {\bf 2019}, {\em 3},~755--759.

\bibitem[{Liu} et~al.(2012){Liu}, {Orosz} and {Bregman}]{2012ApJ...745...89L}
{Liu}, J.; {Orosz}, J.; {Bregman}, J.N.
\newblock {Dynamical Mass Constraints on the Ultraluminous X-Ray Source NGC
  1313 X-2}.
\newblock {\em  Astrophys. J.} {\bf 2012}, {\em 745},~89.
\newblock {\url{https://doi.org/10.1088/0004-637X/745/1/89}}.

\bibitem[{Paynter} et~al.(2021){Paynter}, {Webster} and
  {Thrane}]{2021NatAs...5..560P}
{Paynter}, J.; {Webster}, R.; {Thrane}, E.
\newblock {Evidence for an intermediate-mass black hole from a gravitationally
  lensed gamma-ray burst}.
\newblock {\em Nat. Astron.} {\bf 2021}, {\em 5},~560--568.

\bibitem[{Greene} and {Ho}(2004)]{2004ApJ...610..722G}
{Greene}, J.E.; {Ho}, L.C.
\newblock {Active Galactic Nuclei with Candidate Intermediate-Mass Black
  Holes}.
\newblock {\em  Astrophys. J.} {\bf 2004}, {\em 610},~722--736.

\bibitem[{Greene} and {Ho}(2007)]{2007ApJ...670...92G}
{Greene}, J.E.; {Ho}, L.C.
\newblock {A New Sample of Low-Mass Black Holes in Active Galaxies}.
\newblock {\em  Astrophys. J.} {\bf 2007}, {\em 670},~92--104.

\bibitem[{Dong} et~al.(2012){Dong}, {Ho}, {Yuan}, {Wang}, {Fan}, {Zhou} and
  {Jiang}]{2012ApJ...755..167D}
{Dong}, X.B.; {Ho}, L.C.; {Yuan}, W.; {Wang}, T.G.; {Fan}, X.; {Zhou}, H.;
  {Jiang}, N.
\newblock {A Uniformly Selected Sample of Low-mass Black Holes in Seyfert 1
  Galaxies}.
\newblock {\em  Astrophys. J.} {\bf 2012}, {\em 755},~167.

\bibitem[{Reines} et~al.(2013){Reines}, {Greene} and
  {Geha}]{2013ApJ...775..116R}
{Reines}, A.E.; {Greene}, J.E.; {Geha}, M.
\newblock {Dwarf Galaxies with Optical Signatures of Active Massive Black
  Holes}.
\newblock {\em  Astrophys. J.} {\bf 2013}, {\em 775},~116.

\bibitem[{Abramowicz} et~al.(2004){Abramowicz}, {Klu{\'z}niak}, {McClintock},
  and {Remillard}]{2004ApJ...609L..63A}
{Abramowicz}, M.A.; {Klu{\'z}niak}, W.; {McClintock}, J.E.; {Remillard}, R.A.
\newblock {The Importance of Discovering a 3:2 Twin-Peak Quasi-periodic
  Oscillation in an Ultraluminous X-Ray Source or How to Solve the Puzzle of
  Intermediate-Mass Black Holes}.
\newblock {\em  Astrophys. J.} {\bf 2004}, {\em 609},~L63--L65.
\newblock {\url{https://doi.org/10.1086/422810}}.

\bibitem[{Abramowicz} and {Liu}(2012)]{2012A&A...548A...3A}
{Abramowicz}, M.A.; {Liu}, F.K.
\newblock {Mass estimate of the Swift J 164449.3+573451 supermassive black hole
  based on the 3:2 QPO resonance hypothesis}.
\newblock {\em Astron. Astrophys.} {\bf 2012}, {\em 548},~A3.
\newblock {\url{https://doi.org/10.1051/0004-6361/201220254}}.

\bibitem[{Pasham} et~al.(2014){Pasham}, {Strohmayer} and
  {Mushotzky}]{2014Natur.513...74P}
{Pasham}, D.R.; {Strohmayer}, T.E.; {Mushotzky}, R.F.
\newblock {A 400-solar-mass black hole in the galaxy M82}.
\newblock {\em Nature} {\bf 2014}, {\em 513},~74--76.
\newblock {\url{https://doi.org/10.1038/nature13710}}.

\bibitem[{Pasham} et~al.(2015){Pasham}, {Cenko}, {Zoghbi}, {Mushotzky},
  {Miller} and {Tombesi}]{2015ApJ...811L..11P}
{Pasham}, D.R.; {Cenko}, S.B.; {Zoghbi}, A.; {Mushotzky}, R.F.; {Miller}, J.;
  {Tombesi}, F.
\newblock {Evidence for High-frequency QPOs with a 3:2 Frequency Ratio from a
  5000 Solar Mass Black Hole}.
\newblock {\em  Astrophys. J.} {\bf 2015}, {\em 811},~L11.

\bibitem[{Goluchov{\'a}} et~al.(2019){Goluchov{\'a}}, {T{\"o}r{\"o}k},
  {{\v{S}}r{\'a}mkov{\'a}}, {Abramowicz}, {Stuchl{\'\i}k} and
  {Hor{\'a}k}]{2019A&A...622L...8G}
{Goluchov{\'a}}, K.; {T{\"o}r{\"o}k}, G.; {{\v{S}}r{\'a}mkov{\'a}}, E.;
  {Abramowicz}, M.A.; {Stuchl{\'\i}k}, Z.; {Hor{\'a}k}, J.
\newblock {Mass of the active galactic nucleus black hole
  XMMUJ134736.6+173403}.
\newblock {\em Astron. Astrophys.} {\bf 2019}, {\em 622},~L8.
\newblock {\url{https://doi.org/10.1051/0004-6361/201834774}}.

\bibitem[{Portegies Zwart} et~al.(2004){Portegies Zwart}, {Dewi} and
  {Maccarone}]{2004MNRAS.355..413P}
{Portegies Zwart}, S.F.; {Dewi}, J.; {Maccarone}, T.
\newblock {Intermediate mass black holes in accreting binaries: Formation,
  evolution and observational appearance}.
\newblock {\em Mon. Not. R. Astron. Soc.} {\bf 2004}, {\em 355},~413--423.
\newblock {\url{https://doi.org/10.1111/j.1365-2966.2004.08327.x}}.

\bibitem[{Webb} et~al.(2012){Webb}, {Cseh}, {Lenc}, {Godet}, {Barret},
  {Corbel}, {Farrell}, {Fender}, {Gehrels} and {Heywood}]{2012Sci...337..554W}
{Webb}, N.; {Cseh}, D.; {Lenc}, E.; {Godet}, O.; {Barret}, D.; {Corbel}, S.;
  {Farrell}, S.; {Fender}, R.; {Gehrels}, N.; {Heywood}, I.
\newblock {Radio Detections During Two State Transitions of the
  Intermediate-Mass Black Hole HLX-1}.
\newblock {\em Science} {\bf 2012}, {\em 337},~554.

\bibitem[{Kaaret} et~al.(2017){Kaaret}, {Feng} and
  {Roberts}]{2017ARA&A..55..303K}
{Kaaret}, P.; {Feng}, H.; {Roberts}, T.P.
\newblock {Ultraluminous X-Ray Sources}.
\newblock {\em Annu. Rev. Astron. Astrophys.} {\bf 2017}, {\em 55},~303--341.
\newblock {\url{https://doi.org/10.1146/annurev-astro-091916-055259}}.

\bibitem[{Gebhardt} et~al.(2005){Gebhardt}, {Rich} and
  {Ho}]{2005ApJ...634.1093G}
{Gebhardt}, K.; {Rich}, R.M.; {Ho}, L.C.
\newblock {An Intermediate-Mass Black Hole in the Globular Cluster G1: Improved
  Significance from New Keck and Hubble Space Telescope Observations}.
\newblock {\em  Astrophys. J.} {\bf 2005}, {\em 634},~1093--1102.

\bibitem[{K{\i}z{\i}ltan} et~al.(2017){K{\i}z{\i}ltan}, {Baumgardt} and
  {Loeb}]{2017Natur.542..203K}
{K{\i}z{\i}ltan}, B.; {Baumgardt}, H.; {Loeb}, A.
\newblock {An intermediate-mass black hole in the centre of the globular
  cluster 47 Tucanae}.
\newblock {\em Nature} {\bf 2017}, {\em 542},~203--205.

\bibitem[{Wen} et~al.(2021){Wen}, {Jonker}, {Stone} and
  {Zabludoff}]{2021ApJ...918...46W}
{Wen}, S.; {Jonker}, P.G.; {Stone}, N.C.; {Zabludoff}, A.I.
\newblock {Mass, Spin and Ultralight Boson Constraints from the
  Intermediate-mass Black Hole in the Tidal Disruption Event 3XMM
  J215022.4-055108}.
\newblock {\em  Astrophys. J.} {\bf 2021}, {\em 918},~46.

\bibitem[{Pechetti} et~al.(2022){Pechetti}, {Seth}, {Kamann}, {Caldwell},
  {Strader}, {den Brok}, {Luetzgendorf}, {Neumayer} and
  {Voggel}]{2022ApJ...924...48P}
{Pechetti}, R.; {Seth}, A.; {Kamann}, S.; {Caldwell}, N.; {Strader}, J.; {den
  Brok}, M.; {Luetzgendorf}, N.; {Neumayer}, N.; {Voggel}, K.
\newblock {Detection of a 100,000 M $_{{\ensuremath{\odot}}}$ black hole in
  M31's Most Massive Globular Cluster: A Tidally Stripped Nucleus}.
\newblock {\em  Astrophys. J.} {\bf 2022}, {\em 924},~48.

\bibitem[{Filippenko} and {Ho}(2003)]{2003ApJ...588L..13F}
{Filippenko}, A.V.; {Ho}, L.C.
\newblock {A Low-Mass Central Black Hole in the Bulgeless Seyfert 1 Galaxy NGC
  4395}.
\newblock {\em  Astrophys. J.} {\bf 2003}, {\em 588},~L13--L16.

\bibitem[{Liu} et~al.(2018){Liu}, {Yuan}, {Dong}, {Zhou} and
  {Liu}]{2018ApJS..235...40L}
{Liu}, H.Y.; {Yuan}, W.; {Dong}, X.B.; {Zhou}, H.; {Liu}, W.J.
\newblock {A Uniformly Selected Sample of Low-mass Black Holes in Seyfert 1
  Galaxies. II. The SDSS DR7 Sample}.
\newblock {\em  Astrophys. J.} {\bf 2018}, {\em 235},~40.

\bibitem[{Chilingarian} et~al.(2018){Chilingarian}, {Katkov}, {Zolotukhin},
  {Grishin}, {Beletsky}, {Boutsia} and {Osip}]{2018ApJ...863....1C}
{Chilingarian}, I.V.; {Katkov}, I.Y.; {Zolotukhin}, I.Y.; {Grishin}, K.A.;
  {Beletsky}, Y.; {Boutsia}, K.; {Osip}, D.J.
\newblock {A Population of Bona Fide Intermediate-mass Black Holes Identified
  as Low-luminosity Active Galactic Nuclei}.
\newblock {\em  Astrophys. J.} {\bf 2018}, {\em 863},~1.

\bibitem[{Greene}(2012)]{2012NatCo...3.1304G}
{Greene}, J.E.
\newblock {Low-mass black holes as the remnants of primordial black hole
  formation}.
\newblock {\em Nat. Commun.} {\bf 2012}, {\em 3},~1304.
\newblock {\url{https://doi.org/10.1038/ncomms2314}}.

\bibitem[{Falcke} and {Biermann}(1995)]{1995A&A...293..665F}
{Falcke}, H.; {Biermann}, P.L.
\newblock {The jet--disk symbiosis. I. Radio to X-ray emission models for
  quasars.}
\newblock {\em Astron. Astrophys.} {\bf 1995}, {\em 293},~665--682.

\bibitem[{Merloni} et~al.(2003){Merloni}, {Heinz} and {di
  Matteo}]{2003MNRAS.345.1057M}
{Merloni}, A.; {Heinz}, S.; {di Matteo}, T.
\newblock {A Fundamental Plane of black hole activity}.
\newblock {\em Mon. Not. R. Astron. Soc.} {\bf 2003}, {\em 345},~1057--1076.

\bibitem[{G{\"u}ltekin} et~al.(2014){G{\"u}ltekin}, {Cackett}, {King},
  {Miller} and {Pinkney}]{2014ApJ...788L..22G}
{G{\"u}ltekin}, K.; {Cackett}, E.M.; {King}, A.L.; {Miller}, J.M.; {Pinkney},
  J.
\newblock {Low-mass AGNs and Their Relation to the Fundamental Plane of Black
  Hole Accretion}.
\newblock {\em  Astrophys. J.} {\bf 2014}, {\em 788},~L22.

\bibitem[{Greene} et~al.(2006){Greene}, {Ho} and
  {Ulvestad}]{2006ApJ...636...56G}
{Greene}, J.E.; {Ho}, L.C.; {Ulvestad}, J.S.
\newblock {The Radio Quiescence of Active Galaxies with High Accretion Rates}.
\newblock {\em  Astrophys. J.} {\bf 2006}, {\em 636},~56--62.

\bibitem[{McClintock} and {Remillard}(2006)]{2006csxs.book..157M}
{McClintock}, J.E.; {Remillard}, R.A.
\newblock {Black hole binaries}. \newblock {\em  Cambridge Astrophysics Series, No. 39. Cambridge, UK: Cambridge University Press.} {\bf 2006};
  Volume~39, pp. 157--213.

\bibitem[{Yang} et~al.(2020){Yang}, {Yao}, {Yang}, {Ho}, {An}, {Wang}, {Baan},
  {Gu}, {Liu}, {Yang} and {Joshi}]{2020ApJ...904..200Y}
{Yang}, X.; {Yao}, S.; {Yang}, J.; {Ho}, L.C.; {An}, T.; {Wang}, R.; {Baan},
  W.A.; {Gu}, M.; {Liu}, X.; {Yang}, X.;  et~al.
\newblock {Radio Activity of Supermassive Black Holes with Extremely High
  Accretion Rates}.
\newblock {\em  Astrophys. J.} {\bf 2020}, {\em 904},~200.

\bibitem[{Gallo} et~al.(2003){Gallo}, {Fender} and
  {Pooley}]{2003MNRAS.344...60G}
{Gallo}, E.; {Fender}, R.P.; {Pooley}, G.G.
\newblock {A universal radio-X-ray correlation in low/hard state black hole
  binaries}.
\newblock {\em Mon. Not. R. Astron. Soc.} {\bf 2003}, {\em 344},~60--72.

\bibitem[{Bariuan} et~al.(2022){Bariuan}, {Snios}, {Sobolewska},
  {Siemiginowska} and {Schwartz}]{2022MNRAS.513.4673B}
{Bariuan}, L.G.C.; {Snios}, B.; {Sobolewska}, M.; {Siemiginowska}, A.;
  {Schwartz}, D.A.
\newblock {The Fundamental Planes of black hole activity for radio-loud and
  radio-quiet quasars}.
\newblock {\em Mon. Not. R. Astron. Soc.} {\bf 2022}, {\em 513},~4673--4681.

\bibitem[{Yang} et~al.(2022){Yang}, {Mohan}, {Yang}, {Ho}, {Aditya}, {Zhang},
  {Jaiswal} and {Yang}]{yang2022radio}
{Yang}, X.; {Mohan}, P.; {Yang}, J.; {Ho}, L.C.; {Aditya}, J.N.H.S.; {Zhang},
  S.; {Jaiswal}, S.; {Yang}, X.
\newblock {Radio Observations of Four Active Galactic Nuclei Hosting
  Intermediate-mass Black Hole Candidates: Studying the Outflow Activity and
  Evolution}.
\newblock {\em  Astrophys. J.} {\bf 2022}, {\em 941},~43.
\newblock {\url{https://doi.org/10.3847/1538-4357/ac9e9d}}.

\bibitem[{Blandford} and {K{\"o}nigl}(1979)]{1979ApJ...232...34B}
{Blandford}, R.D.; {K{\"o}nigl}, A.
\newblock {Relativistic jets as compact radio sources.}
\newblock {\em  Astrophys. J.} {\bf 1979}, {\em 232},~34--48.

\bibitem[{Blandford} et~al.(2019){Blandford}, {Meier} and
  {Readhead}]{2019ARA&A..57..467B}
{Blandford}, R.; {Meier}, D.; {Readhead}, A.
\newblock {Relativistic Jets from Active Galactic Nuclei}.
\newblock {\em Annu. Rev. Astron. Astrophys.} {\bf 2019}, {\em 57},~467--509.

\bibitem[{Ba{\~n}ados} et~al.(2015){Ba{\~n}ados}, {Venemans}, {Morganson},
  {Hodge}, {Decarli}, {Walter}, {Stern}, {Schlafly}, {Farina}, {Greiner},
  {Chambers}, {Fan}, {Rix}, {Burgett}, {Draper}, {Flewelling}, {Kaiser},
  {Metcalfe}, {Morgan}, {Tonry} and {Wainscoat}]{2015ApJ...804..118B}
{Ba{\~n}ados}, E.; {Venemans}, B.P.; {Morganson}, E.; {Hodge}, J.; {Decarli},
  R.; {Walter}, F.; {Stern}, D.; {Schlafly}, E.; {Farina}, E.P.; {Greiner}, J.;
   et~al.
\newblock {Constraining the Radio-loud Fraction of Quasars at z > 5.5}.
\newblock {\em  Astrophys. J.} {\bf 2015}, {\em 804},~118.
\newblock {\url{https://doi.org/10.1088/0004-637X/804/2/118}}.

\bibitem[{Panessa} et~al.(2019){Panessa}, {Baldi}, {Laor}, {Padovani}, {Behar},
  and {McHardy}]{2019NatAs...3..387P}
{Panessa}, F.; {Baldi}, R.D.; {Laor}, A.; {Padovani}, P.; {Behar}, E.;
  {McHardy}, I.
\newblock {The origin of radio emission from radio-quiet active galactic
  nuclei}.
\newblock {\em Nat. Astron.} {\bf 2019}, {\em 3},~387--396.

\bibitem[{Yang} et~al.(2020){Yang}, {Paragi}, {An}, {Baan}, {Mohan} and
  {Liu}]{2020MNRAS.494.1744Y}
{Yang}, J.; {Paragi}, Z.; {An}, T.; {Baan}, W.A.; {Mohan}, P.; {Liu}, X.
\newblock {A two-sided but significantly beamed jet in the supercritical
  accretion quasar IRAS F11119+3257}.
\newblock {\em Mon. Not. R. Astron. Soc.} {\bf 2020}, {\em 494},~1744--1750.

\bibitem[{Yang} et~al.(2021){Yang}, {Paragi}, {Nardini}, {Baan}, {Fan},
  {Mohan}, {Varenius} and {An}]{2021MNRAS.500.2620Y}
{Yang}, J.; {Paragi}, Z.; {Nardini}, E.; {Baan}, W.A.; {Fan}, L.; {Mohan}, P.;
  {Varenius}, E.; {An}, T.
\newblock {The nearby extreme accretion and feedback system PDS 456: Finding a
  complex radio-emitting nucleus}.
\newblock {\em Mon. Not. R. Astron. Soc.} {\bf 2021}, {\em 500},~2620--2626.

\bibitem[{Wang} et~al.(2021){Wang}, {An}, {Jaiswal}, {Mohan}, {Wang}, {Baan},
  {Zhang} and {Yang}]{2021MNRAS.504.3823W}
{Wang}, A.; {An}, T.; {Jaiswal}, S.; {Mohan}, P.; {Wang}, Y.; {Baan}, W.A.;
  {Zhang}, Y.; {Yang}, X.
\newblock {The obstructed jet in Mrk 231}.
\newblock {\em Mon. Not. R. Astron. Soc.} {\bf 2021}, {\em 504},~3823--3830.

\bibitem[{Baskin} and {Laor}(2021)]{2021MNRAS.508..680B}
{Baskin}, A.; {Laor}, A.
\newblock {Radiation pressure confinement---V. The predicted free-free
  absorption and emission in active galactic nuclei}.
\newblock {\em Mon. Not. R. Astron. Soc.} {\bf 2021}, {\em 508},~680--697.

\bibitem[{Laor} and {Behar}(2008)]{2008MNRAS.390..847L}
{Laor}, A.; {Behar}, E.
\newblock {On the origin of radio emission in radio-quiet quasars}.
\newblock {\em Mon. Not. R. Astron. Soc.} {\bf 2008}, {\em 390},~847--862.

\bibitem[{Raginski} and {Laor}(2016)]{2016MNRAS.459.2082R}
{Raginski}, I.; {Laor}, A.
\newblock {AGN coronal emission models---I. The predicted radio emission}.
\newblock {\em Mon. Not. R. Astron. Soc.} {\bf 2016}, {\em 459},~2082--2096.

\bibitem[{Inoue} and {Doi}(2018)]{2018ApJ...869..114I}
{Inoue}, Y.; {Doi}, A.
\newblock {Detection of Coronal Magnetic Activity in nearby Active Supermassive
  Black Holes}.
\newblock {\em  Astrophys. J.} {\bf 2018}, {\em 869},~114.

\bibitem[{Zakamska} and {Greene}(2014)]{2014MNRAS.442..784Z}
{Zakamska}, N.L.; {Greene}, J.E.
\newblock {Quasar feedback and the origin of radio emission in radio-quiet
  quasars}.
\newblock {\em Mon. Not. R. Astron. Soc.} {\bf 2014}, {\em 442},~784--804.

\bibitem[{Nims} et~al.(2015){Nims}, {Quataert} and
  {Faucher-Gigu{\`e}re}]{2015MNRAS.447.3612N}
{Nims}, J.; {Quataert}, E.; {Faucher-Gigu{\`e}re}, C.A.
\newblock {Observational signatures of galactic winds powered by active
  galactic nuclei}.
\newblock {\em Mon. Not. R. Astron. Soc.} {\bf 2015}, {\em 447},~3612--3622.

\bibitem[{Wrobel} and {Ho}(2006)]{2006ApJ...646L..95W}
{Wrobel}, J.M.; {Ho}, L.C.
\newblock {Radio Emission on Subparsec Scales from the Intermediate-Mass Black
  Hole in NGC 4395}.
\newblock {\em  Astrophys. J.} {\bf 2006}, {\em 646},~L95--L98.

\bibitem[{Reines} and {Deller}(2012)]{2012ApJ...750L..24R}
{Reines}, A.E.; {Deller}, A.T.
\newblock {Parsec-scale Radio Emission from the Low-luminosity Active Galactic
  Nucleus in the Dwarf Starburst Galaxy Henize 2--10}.
\newblock {\em  Astrophys. J.} {\bf 2012}, {\em 750},~L24.

\bibitem[{Paragi} et~al.(2014){Paragi}, {Frey}, {Kaaret}, {Cseh}, {Overzier},
  and {Kharb}]{2014ApJ...791....2P}
{Paragi}, Z.; {Frey}, S.; {Kaaret}, P.; {Cseh}, D.; {Overzier}, R.; {Kharb}, P.
\newblock {Probing the Active Massive Black Hole Candidate in the Center of NGC
  404 with VLBI}.
\newblock {\em  Astrophys. J.} {\bf 2014}, {\em 791},~2.

\bibitem[{Yang} et~al.(2020){Yang}, {Gurvits}, {Paragi}, {Frey}, {Conway},
  {Liu} and {Cui}]{2020MNRAS.495L..71Y}
{Yang}, J.; {Gurvits}, L.I.; {Paragi}, Z.; {Frey}, S.; {Conway}, J.E.; {Liu},
  X.; {Cui}, L.
\newblock {A parsec-scale radio jet launched by the central intermediate-mass
  black hole in the dwarf galaxy SDSS J090613.77+561015.2}.
\newblock {\em Mon. Not. R. Astron. Soc.} {\bf 2020}, {\em 495},~L71--L75.

\bibitem[{Terashima} and {Wilson}(2003)]{2003ApJ...583..145T}
{Terashima}, Y.; {Wilson}, A.S.
\newblock {Chandra Snapshot Observations of Low-Luminosity Active Galactic
  Nuclei with a Compact Radio Source}.
\newblock {\em  Astrophys. J.} {\bf 2003}, {\em 583},~145--158.

\bibitem[{Yang} et~al.(2022){Yang}, {Yang}, {Wrobel}, {Paragi}, {Gurvits},
  {Ho}, {Nyland}, {Fan} and {Tafoya}]{2022MNRAS.514.6215Y}
{Yang}, J.; {Yang}, X.; {Wrobel}, J.M.; {Paragi}, Z.; {Gurvits}, L.I.; {Ho},
  L.C.; {Nyland}, K.; {Fan}, L.; {Tafoya}, D.
\newblock {Is there a sub-parsec-scale jet base in the nearby dwarf galaxy NGC
  4395?}
\newblock {\em Mon. Not. R. Astron. Soc.} {\bf 2022}, {\em 514},~6215--6224.

\bibitem[{Nyland} et~al.(2012){Nyland}, {Marvil}, {Wrobel}, {Young} and
  {Zauderer}]{2012ApJ...753..103N}
{Nyland}, K.; {Marvil}, J.; {Wrobel}, J.M.; {Young}, L.M.; {Zauderer}, B.A.
\newblock {The Intermediate-mass Black Hole Candidate in the Center of NGC 404:
  New Evidence from Radio Continuum Observations}.
\newblock {\em  Astrophys. J.} {\bf 2012}, {\em 753},~103.

\bibitem[{Esin} et~al.(1997){Esin}, {McClintock} and
  {Narayan}]{1997ApJ...489..865E}
{Esin}, A.A.; {McClintock}, J.E.; {Narayan}, R.
\newblock {Advection-Dominated Accretion and the Spectral States of Black Hole
  X-Ray Binaries: Application to Nova Muscae 1991}.
\newblock {\em  Astrophys. J.} {\bf 1997}, {\em 489},~865--889.
\newblock {\url{https://doi.org/10.1086/304829}}.

\bibitem[{Romero} et~al.(2017){Romero}, {Boettcher}, {Markoff} and
  {Tavecchio}]{2017SSRv..207....5R}
{Romero}, G.E.; {Boettcher}, M.; {Markoff}, S.; {Tavecchio}, F.
\newblock {Relativistic Jets in Active Galactic Nuclei and Microquasars}.
\newblock {\em Space Sci. Rev.} {\bf 2017}, {\em 207},~5--61.

\bibitem[{Fender} et~al.(2004){Fender}, {Belloni} and
  {Gallo}]{2004MNRAS.355.1105F}
{Fender}, R.P.; {Belloni}, T.M.; {Gallo}, E.
\newblock {Towards a unified model for black hole X-ray binary jets}.
\newblock {\em Mon. Not. R. Astron. Soc.} {\bf 2004}, {\em 355},~1105--1118.

\bibitem[{Ruan} et~al.(2019){Ruan}, {Anderson}, {Eracleous}, {Green},
  {Haggard}, {MacLeod}, {Runnoe} and {Sobolewska}]{2019ApJ...883...76R}
{Ruan}, J.J.; {Anderson}, S.F.; {Eracleous}, M.; {Green}, P.J.; {Haggard}, D.;
  {MacLeod}, C.L.; {Runnoe}, J.C.; {Sobolewska}, M.A.
\newblock {The Analogous Structure of Accretion Flows in Supermassive and
  Stellar Mass Black Holes: New Insights from Faded Changing-look Quasars}.
\newblock {\em  Astrophys. J.} {\bf 2019}, {\em 883},~76.
\newblock {\url{https://doi.org/10.3847/1538-4357/ab3c1a}}.

\bibitem[{Blandford} and {Payne}(1982)]{1982MNRAS.199..883B}
{Blandford}, R.D.; {Payne}, D.G.
\newblock {Hydromagnetic flows from accretion disks and the production of radio
  jets.}
\newblock {\em Mon. Not. R. Astron. Soc.} {\bf 1982}, {\em 199},~883--903.
\newblock {\url{https://doi.org/10.1093/mnras/199.4.883}}.

\bibitem[{Blandford} and {Znajek}(1977)]{1977MNRAS.179..433B}
{Blandford}, R.D.; {Znajek}, R.L.
\newblock {Electromagnetic extraction of energy from Kerr black holes.}
\newblock {\em Mon. Not. R. Astron. Soc.} {\bf 1977}, {\em 179},~433--456.
\newblock {\url{https://doi.org/10.1093/mnras/179.3.433}}.

\bibitem[{Shende} et~al.(2019){Shende}, {Subramanian} and
  {Sachdeva}]{2019ApJ...877..130S}
{Shende}, M.B.; {Subramanian}, P.; {Sachdeva}, N.
\newblock {Episodic Jets from Black Hole Accretion Disks}.
\newblock {\em  Astrophys. J.} {\bf 2019}, {\em 877},~130.
\newblock {\url{https://doi.org/10.3847/1538-4357/ab1cb6}}.

\bibitem[{Vierdayanti} et~al.(2013){Vierdayanti}, {Sadowski}, {Mineshige} and
  {Bursa}]{2013MNRAS.436...71V}
{Vierdayanti}, K.; {Sadowski}, A.; {Mineshige}, S.; {Bursa}, M.
\newblock {Inner disc obscuration in GRS 1915+105 based on relativistic slim
  disc model}.
\newblock {\em Mon. Not. R. Astron. Soc.} {\bf 2013}, {\em 436},~71--81.
\newblock {\url{https://doi.org/10.1093/mnras/stt1467}}.

\bibitem[{Gladstone} et~al.(2009){Gladstone}, {Roberts} and
  {Done}]{2009MNRAS.397.1836G}
{Gladstone}, J.C.; {Roberts}, T.P.; {Done}, C.
\newblock {The ultraluminous state}.
\newblock {\em Mon. Not. R. Astron. Soc.} {\bf 2009}, {\em 397},~1836--1851.
\newblock {\url{https://doi.org/10.1111/j.1365-2966.2009.15123.x}}.

\bibitem[{Svoboda} et~al.(2017){Svoboda}, {Guainazzi} and
  {Merloni}]{2017A&A...603A.127S}
{Svoboda}, J.; {Guainazzi}, M.; {Merloni}, A.
\newblock {AGN spectral states from simultaneous UV and X-ray observations by
  XMM-Newton}.
\newblock {\em Astron. Astrophys.} {\bf 2017}, {\em 603},~A127.

\bibitem[{Falcke} et~al.(2004){Falcke}, {K{\"o}rding} and
  {Markoff}]{2004A&A...414..895F}
{Falcke}, H.; {K{\"o}rding}, E.; {Markoff}, S.
\newblock {A scheme to unify low-power accreting black holes. Jet-dominated
  accretion flows and the radio/X-ray correlation}.
\newblock {\em Astron. Astrophys.} {\bf 2004}, {\em 414},~895--903.

\bibitem[{Ho}(2002)]{2002ApJ...564..120H}
{Ho}, L.C.
\newblock {On the Relationship between Radio Emission and Black Hole Mass in
  Galactic Nuclei}.
\newblock {\em  Astrophys. J.} {\bf 2002}, {\em 564},~120--132.

\bibitem[{Sikora} et~al.(2007){Sikora}, {Stawarz} and
  {Lasota}]{2007ApJ...658..815S}
{Sikora}, M.; {Stawarz}, {\L}.; {Lasota}, J.P.
\newblock {Radio Loudness of Active Galactic Nuclei: Observational Facts and
  Theoretical Implications}.
\newblock {\em  Astrophys. J.} {\bf 2007}, {\em 658},~815--828.

\bibitem[{Broderick} and {Fender}(2011)]{2011MNRAS.417..184B}
{Broderick}, J.W.; {Fender}, R.P.
\newblock {Is there really a dichotomy in active galactic nucleus jet power?}
\newblock {\em Mon. Not. R. Astron. Soc.} {\bf 2011}, {\em 417},~184--197.

\bibitem[{McHardy} et~al.(2006){McHardy}, {Koerding}, {Knigge}, {Uttley} and
  {Fender}]{2006Natur.444..730M}
{McHardy}, I.M.; {Koerding}, E.; {Knigge}, C.; {Uttley}, P.; {Fender}, R.P.
\newblock {Active galactic nuclei as scaled-up Galactic black holes}.
\newblock {\em Nature} {\bf 2006}, {\em 444},~730--732.
\newblock {\url{https://doi.org/10.1038/nature05389}}.

\bibitem[{K{\"o}rding} et~al.(2006){K{\"o}rding}, {Jester} and
  {Fender}]{2006MNRAS.372.1366K}
{K{\"o}rding}, E.G.; {Jester}, S.; {Fender}, R.
\newblock {Accretion states and radio loudness in active galactic nuclei:
  analogies with X-ray binaries}.
\newblock {\em Mon. Not. R. Astron. Soc.} {\bf 2006}, {\em 372},~1366--1378.
\newblock {\url{https://doi.org/10.1111/j.1365-2966.2006.10954.x}}.

\bibitem[{Mondal} et~al.(2014){Mondal}, {Debnath} and
  {Chakrabarti}]{2014ApJ...786....4M}
{Mondal}, S.; {Debnath}, D.; {Chakrabarti}, S.K.
\newblock {Inference on Accretion Flow Dynamics Using TCAF Solution from the
  Analysis of Spectral Evolution of H 1743-322 during the 2010 Outburst}.
\newblock {\em  Astrophys. J.} {\bf 2014}, {\em 786},~4.
\newblock {\url{https://doi.org/10.1088/0004-637X/786/1/4}}.

\bibitem[{Jana} et~al.(2016){Jana}, {Debnath}, {Chakrabarti}, {Mondal} and
  {Molla}]{2016ApJ...819..107J}
{Jana}, A.; {Debnath}, D.; {Chakrabarti}, S.K.; {Mondal}, S.; {Molla}, A.A.
\newblock {Accretion Flow Dynamics of MAXI J1836-194 During Its 2011 Outburst
  from TCAF Solution}.
\newblock {\em  Astrophys. J.} {\bf 2016}, {\em 819},~107.
\newblock {\url{https://doi.org/10.3847/0004-637X/819/2/107}}.

\bibitem[{Zhou} et~al.(2015){Zhou}, {Yuan}, {Pan} and
  {Liu}]{2015ApJ...798L...5Z}
{Zhou}, X.L.; {Yuan}, W.; {Pan}, H.W.; {Liu}, Z.
\newblock {Universal Scaling of the 3:2 Twin-peak Quasi-periodic Oscillation
  Frequencies With Black Hole Mass and Spin Revisited}.
\newblock {\em  Astrophys. J.} {\bf 2015}, {\em 798},~L5.
\newblock {\url{https://doi.org/10.1088/2041-8205/798/1/L5}}.

\bibitem[{Reines} et~al.(2020){Reines}, {Condon}, {Darling} and
  {Greene}]{2020ApJ...888...36R}
{Reines}, A.E.; {Condon}, J.J.; {Darling}, J.; {Greene}, J.E.
\newblock {A New Sample of (Wandering) Massive Black Holes in Dwarf Galaxies
  from High-resolution Radio Observations}.
\newblock {\em  Astrophys. J.} {\bf 2020}, {\em 888},~36.

\bibitem[{Condon} et~al.(1998){Condon}, {Cotton}, {Greisen}, {Yin}, {Perley},
  {Taylor} and {Broderick}]{1998AJ....115.1693C}
{Condon}, J.J.; {Cotton}, W.D.; {Greisen}, E.W.; {Yin}, Q.F.; {Perley}, R.A.;
  {Taylor}, G.B.; {Broderick}, J.J.
\newblock {The NRAO VLA Sky Survey}.
\newblock {\em  Astron. J.} {\bf 1998}, {\em 115},~1693--1716.

\bibitem[{Becker} et~al.(1995){Becker}, {White} and
  {Helfand}]{1995ApJ...450..559B}
{Becker}, R.H.; {White}, R.L.; {Helfand}, D.J.
\newblock {The FIRST Survey: Faint Images of the Radio Sky at Twenty
  Centimeters}.
\newblock {\em  Astrophys. J.} {\bf 1995}, {\em 450},~559.

\bibitem[{Davis} et~al.(2020){Davis}, {Nguyen}, {Seth}, {Greene}, {Nyland},
  {Barth}, {Bureau}, {Cappellari}, {den Brok}, {Iguchi}, {Lelli}, {Liu},
  {Neumayer}, {North}, {Onishi}, {Sarzi}, {Smith} and
  {Williams}]{2020MNRAS.496.4061D}
{Davis}, T.A.; {Nguyen}, D.D.; {Seth}, A.C.; {Greene}, J.E.; {Nyland}, K.;
  {Barth}, A.J.; {Bureau}, M.; {Cappellari}, M.; {den Brok}, M.; {Iguchi}, S.;
  et~al.
\newblock {Revealing the intermediate-mass black hole at the heart of the dwarf
  galaxy NGC 404 with sub-parsec resolution ALMA observations}.
\newblock {\em Mon. Not. R. Astron. Soc.} {\bf 2020}, {\em 496},~4061--4078.

\bibitem[{Baldassare} et~al.(2017){Baldassare}, {Reines}, {Gallo} and
  {Greene}]{2017ApJ...836...20B}
{Baldassare}, V.F.; {Reines}, A.E.; {Gallo}, E.; {Greene}, J.E.
\newblock {X-ray and Ultraviolet Properties of AGNs in Nearby Dwarf Galaxies}.
\newblock {\em  Astrophys. J.} {\bf 2017}, {\em 836},~20.
\newblock {\url{https://doi.org/10.3847/1538-4357/836/1/20}}.

\bibitem[{Soria} et~al.(2022){Soria}, {Kolehmainen}, {Graham}, {Swartz},
  {Yukita}, {Motch}, {Jarrett}, {Miller-Jones}, {Plotkin}, {Maccarone},
  {Ferrarese}, {Guest} and {Lan{\c{c}}on}]{2022MNRAS.512.3284S}
{Soria}, R.; {Kolehmainen}, M.; {Graham}, A.W.; {Swartz}, D.A.; {Yukita}, M.;
  {Motch}, C.; {Jarrett}, T.H.; {Miller-Jones}, J.C.A.; {Plotkin}, R.M.;
  {Maccarone}, T.J.;  et~al.
\newblock {A Chandra Virgo cluster survey of spiral galaxies---I. Introduction
  to the survey and a new ULX sample}.
\newblock {\em Mon. Not. R. Astron. Soc.} {\bf 2022}, {\em 512},~3284--3311.
\newblock {\url{https://doi.org/10.1093/mnras/stac148}}.

\bibitem[{Moran} et~al.(2005){Moran}, {Eracleous}, {Leighly}, {Chartas},
  {Filippenko}, {Ho} and {Blanco}]{2005AJ....129.2108M}
{Moran}, E.C.; {Eracleous}, M.; {Leighly}, K.M.; {Chartas}, G.; {Filippenko},
  A.V.; {Ho}, L.C.; {Blanco}, P.R.
\newblock {Extreme X-Ray Behavior of the Low-Luminosity Active Nucleus in NGC
  4395}.
\newblock {\em  Astron. J.} {\bf 2005}, {\em 129},~2108--2118.

\bibitem[{G{\"u}ltekin} et~al.(2022){G{\"u}ltekin}, {Nyland}, {Gray}, {Fehmer},
  {Huang}, {Sparkman}, {Reines}, {Greene}, {Cackett} and
  {Baldassare}]{2022MNRAS.516.6123G}
{G{\"u}ltekin}, K.; {Nyland}, K.; {Gray}, N.; {Fehmer}, G.; {Huang}, T.;
  {Sparkman}, M.; {Reines}, A.E.; {Greene}, J.E.; {Cackett}, E.M.;
  {Baldassare}, V.
\newblock {Intermediate-mass black holes and the Fundamental Plane of black
  hole accretion}.
\newblock {\em Mon. Not. R. Astron. Soc.} {\bf 2022}, {\em 516},~6123--6131.
\newblock {\url{https://doi.org/10.1093/mnras/stac2608}}.

\bibitem[{Saikia} et~al.(2018){Saikia}, {K{\"o}rding}, {Coppejans}, {Falcke},
  {Williams}, {Baldi}, {Mchardy} and {Beswick}]{2018A&A...616A.152S}
{Saikia}, P.; {K{\"o}rding}, E.; {Coppejans}, D.L.; {Falcke}, H.; {Williams},
  D.; {Baldi}, R.D.; {Mchardy}, I.; {Beswick}, R.
\newblock {15-GHz radio emission from nearby low-luminosity active galactic
  nuclei}.
\newblock {\em Astron. Astrophys.} {\bf 2018}, {\em 616},~A152.



\bibitem[{O'Dea} and {Saikia}(2021)]{2021A&ARv..29....3O}
{O'Dea}, C.P.; {Saikia}, D.J.
\newblock {Compact steep-spectrum and peaked-spectrum radio sources}.
\newblock {\em  Astron. Astrophys. Rev.} {\bf 2021}, {\em 29},~3.

\bibitem[{Stawarz} et~al.(2008){Stawarz}, {Ostorero}, {Begelman}, {Moderski},
  {Kataoka} and {Wagner}]{2008ApJ...680..911S}
{Stawarz}, {\L}.; {Ostorero}, L.; {Begelman}, M.C.; {Moderski}, R.; {Kataoka},
  J.; {Wagner}, S.
\newblock {Evolution of and High-Energy Emission from GHz-Peaked Spectrum
  Sources}.
\newblock {\em  Astrophys. J.} {\bf 2008}, {\em 680},~911--925.
\newblock {\url{https://doi.org/10.1086/587781}}.

\bibitem[{G{\"u}ltekin} et~al.(2019){G{\"u}ltekin}, {King}, {Cackett},
  {Nyland}, {Miller}, {Di Matteo}, {Markoff} and {Rupen}]{2019ApJ...871...80G}
{G{\"u}ltekin}, K.; {King}, A.L.; {Cackett}, E.M.; {Nyland}, K.; {Miller},
  J.M.; {Di Matteo}, T.; {Markoff}, S.; {Rupen}, M.P.
\newblock {The Fundamental Plane of Black Hole Accretion and Its Use as a Black
  Hole-Mass Estimator}.
\newblock {\em  Astrophys. J.} {\bf 2019}, {\em 871},~80.
\newblock {\url{https://doi.org/10.3847/1538-4357/aaf6b9}}.

\bibitem[{Fischer} et~al.(2021){Fischer}, {Secrest}, {Johnson}, {Dorland},
  {Cigan}, {Fernandez}, {Hunt}, {Koss}, {Schmitt} and
  {Zacharias}]{2021ApJ...906...88F}
{Fischer}, T.C.; {Secrest}, N.J.; {Johnson}, M.C.; {Dorland}, B.N.; {Cigan},
  P.J.; {Fernandez}, L.C.; {Hunt}, L.R.; {Koss}, M.; {Schmitt}, H.R.;
  {Zacharias}, N.
\newblock {Fundamental Reference AGN Monitoring Experiment (FRAMEx). I. Jumping
  Out of the Plane with the VLBA}.
\newblock {\em  Astrophys. J.} {\bf 2021}, {\em 906},~88.

\bibitem[{Corbel} et~al.(2000){Corbel}, {Fender}, {Tzioumis}, {Nowak},
  {McIntyre}, {Durouchoux} and {Sood}]{2000A&A...359..251C}
{Corbel}, S.; {Fender}, R.P.; {Tzioumis}, A.K.; {Nowak}, M.; {McIntyre}, V.;
  {Durouchoux}, P.; {Sood}, R.
\newblock {Coupling of the X-ray and radio emission in the black hole candidate
  and compact jet source GX 339-4}.
\newblock {\em Astron. Astrophys.} {\bf 2000}, {\em 359},~251--268.
\newblock {\url{https://doi.org/10.48550/arXiv.astro-ph/0003460}}.

\bibitem[{Yang}(2023)]{yang2023}
{Yang}, X.
\newblock {In preparation}. {\bf 2023}.

\bibitem[{Nemmen} et~al.(2007){Nemmen}, {Bower}, {Babul} and
  {Storchi-Bergmann}]{2007MNRAS.377.1652N}
{Nemmen}, R.S.; {Bower}, R.G.; {Babul}, A.; {Storchi-Bergmann}, T.
\newblock {Models for jet power in elliptical galaxies: A case for rapidly
  spinning black holes}.
\newblock {\em Mon. Not. R. Astron. Soc.} {\bf 2007}, {\em 377},~1652--1662.
\newblock {\url{https://doi.org/10.1111/j.1365-2966.2007.11726.x}}.

\bibitem[{Narayan} and {Yi}(1995)]{1995ApJ...452..710N}
{Narayan}, R.; {Yi}, I.
\newblock {Advection-dominated Accretion: Underfed Black Holes and Neutron
  Stars}.
\newblock {\em  Astrophys. J.} {\bf 1995}, {\em 452},~710.
\newblock {\url{https://doi.org/10.1086/176343}}.

\bibitem[{Yuan} et~al.(2009){Yuan}, {Lin}, {Wu} and {Ho}]{2009MNRAS.395.2183Y}
{Yuan}, F.; {Lin}, J.; {Wu}, K.; {Ho}, L.C.
\newblock {A magnetohydrodynamical model for the formation of episodic jets}.
\newblock {\em Mon. Not. R. Astron. Soc.} {\bf 2009}, {\em 395},~2183--2188.
\newblock {\url{https://doi.org/10.1111/j.1365-2966.2009.14673.x}}.

\bibitem[{Nemmen} et~al.(2012){Nemmen}, {Georganopoulos}, {Guiriec}, {Meyer},
  {Gehrels} and {Sambruna}]{2012Sci...338.1445N}
{Nemmen}, R.S.; {Georganopoulos}, M.; {Guiriec}, S.; {Meyer}, E.T.; {Gehrels},
  N.; {Sambruna}, R.M.
\newblock {A Universal Scaling for the Energetics of Relativistic Jets from
  Black Hole Systems}.
\newblock {\em Science} {\bf 2012}, {\em 338},~1445.
\newblock {\url{https://doi.org/10.1126/science.1227416}}.

\bibitem[{McDonald} et~al.(2021){McDonald}, {McNamara}, {Calzadilla}, {Chen},
  {Gaspari}, {Hickox}, {Kara} and {Korchagin}]{2021ApJ...908...85M}
{McDonald}, M.; {McNamara}, B.R.; {Calzadilla}, M.S.; {Chen}, C.T.; {Gaspari},
  M.; {Hickox}, R.C.; {Kara}, E.; {Korchagin}, I.
\newblock {Observational Evidence for Enhanced Black Hole Accretion in Giant
  Elliptical Galaxies}.
\newblock {\em  Astrophys. J.} {\bf 2021}, {\em 908},~85. \newblock {\url{https://doi.org/10.3847/1538-4357/abd47f}}.

\bibitem[{Chen} et~al.(2023){Chen}, {Gu}, {Fan}, {Yu}, {Ding}, {Guo} and
  {Xiong}]{2023MNRAS.519.6199C}
{Chen}, Y.; {Gu}, Q.; {Fan}, J.; {Yu}, X.; {Ding}, N.; {Guo}, X.; {Xiong}, D.
\newblock {Jet power, intrinsic {\ensuremath{\gamma}}-ray luminosity and
  accretion in jetted AGNs}.
\newblock {\em Mon. Not. R. Astron. Soc.} {\bf 2023}, {\em 519},~6199--6209.
\newblock {\url{https://doi.org/10.1093/mnras/stad065}}.

\bibitem[{Chen} et~al.(2021){Chen}, {Gu}, {Fan}, {Zhou}, {Yuan}, {Gu}, {Wu},
  {Xiong}, {Guo}, {Ding} and {Yu}]{2021ApJ...913...93C}
{Chen}, Y.; {Gu}, Q.; {Fan}, J.; {Zhou}, H.; {Yuan}, Y.; {Gu}, W.; {Wu}, Q.;
  {Xiong}, D.; {Guo}, X.; {Ding}, N.;  et~al.
\newblock {The Powers of Relativistic Jets Depend on the Spin of Accreting
  Supermassive Black Holes}.
\newblock {\em  Astrophys. J.} {\bf 2021}, {\em 913},~93. \newblock {\url{https://doi.org/10.3847/1538-4357/abf4ff}}.

\bibitem[{Harrison} et~al.(2018){Harrison}, {Gottlieb} and
  {Nakar}]{2018MNRAS.477.2128H}
{Harrison}, R.; {Gottlieb}, O.; {Nakar}, E.
\newblock {Numerically calibrated model for propagation of a relativistic
  unmagnetized jet in dense media}.
\newblock {\em Mon. Not. R. Astron. Soc.} {\bf 2018}, {\em 477},~2128--2140. \newblock {\url{https://doi.org/10.1093/mnras/sty760}}.

\end{thebibliography}
\end{document}